\documentclass[%
 preprint,
 amsmath,amssymb,
 aps,
prb,
]{revtex4-2}

\usepackage{graphicx}% Include figure files
\usepackage{dcolumn}% Align table columns on decimal point
\usepackage{bm}% bold math
\usepackage{soul,color}% bold math

\begin{document}

\title{How phonon coherence develops and contributes to heat conduction in periodic and aperiodic superlattices}
\author{Theodore Maranets}
\email{tmaranets@unr.edu}
\author{Yan Wang}
\affiliation{Department of Mechanical Engineering, University of Nevada, Reno, Reno, NV, 89557, USA }
\date{\today}

\begin{abstract}
This work investigates the impact of device length on thermal conductivity in periodic and aperiodic superlattices (SLs). While it is well known that thermal conductivity in aperiodic SLs exhibits a weaker dependence on device length compared to periodic SLs, existing literature attributes this behavior to the scattering of coherent phonons by aperiodically arranged interfaces. Through atomistic wave-packet simulations, we show that coherent phonons in aperiodic SLs have spatial extensions limited to a certain number of SL layers, which prevents transmission if the extension is shorter than the device length. Specifically, the disordered interface spacing in aperiodic SLs causes coherent phonons to behave as non-propagative vibrational modes, resulting in diffuse energy transmission. In periodic SLs, however, coherent phonons can propagate across the entire structure, enabling high transmission. The difference between ballistic transport in periodic SLs and diffuse transport in aperiodic SLs is captured in the length-dependence of phonon transmission. These findings provide new insights into phonon coherence and its implications for heat conduction in superlattices, with potential applications in the thermal design of nanostructures.

\end{abstract}

\maketitle

\section{Introduction}

The last decade has seen rising interest in phononic crystals, metamaterials artificially structured to enhance and control the wave properties of phonons \cite{hussein2014dynamics,maldovan2015phonon,volz2016nanophononics,zhang2021coherent,anufriev2021review,ma2023phonon}. A prominent example is superlattices (SLs), which consist of periodically alternating layers of two or more materials. These structures can induce coherent phonon transport when the interface density becomes comparable to phonon wavelengths \cite{Chen_1999,simkin2000minimum,ravichandran2014crossover,latour2014microscopic,zhu2014phonon,xie2018phonon,chen2021non,liu2022heat,zhao2022incoherent}. In the coherent transport regime, interface-scattered phonons interfere constructively to form new resonant modes aligning with the secondary periodicity of the phononic crystal. This phenomenon facilitates free transport across interfaces without scattering, leading to enhanced heat conduction \cite{capinski1999thermal,venkatasubramanian2000lattice,daly2002molecular,latour2017distinguishing,maranets2024prominent}. Modifications to the structure, such as disrupting the secondary periodicity through aperiodic layering, can further tailor the propagation of coherent phonons to ultimately modify the resulting thermophysical properties \cite{wang2014decomposition,wang2015optimization,juntunen2019anderson,hu2021direct,maranets2024influence}. Selective control of phonon wave behaviors makes SLs and similar phononic crystals highly promising for diverse applications, including but not limited to, thermoelectric energy harvesting, thermal management of microelectronic devices, and phase change memory \cite{shi2015evaluating}. 

\begin{figure}
    \centering
    \includegraphics[width=0.5\textwidth]{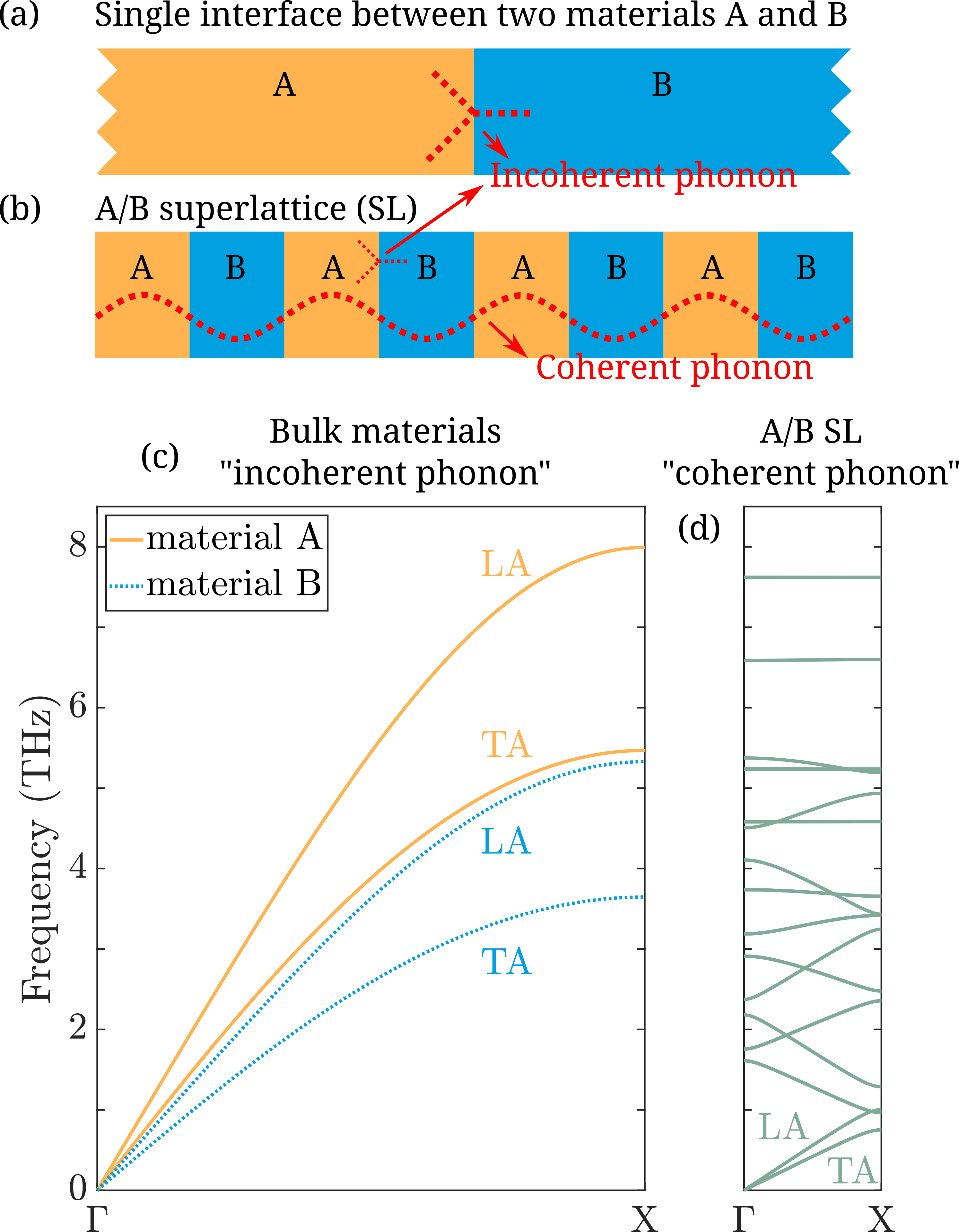}
    \caption{Schematic illustrations of incoherent and coherent phonons within a single interface system (a) and superlattice (b). Incoherent phonons follow the bulk dispersion relation of the base materials (c), while coherent phonons follow the dispersion relation of the superlattice which has secondary periodicity (d). Longitudinal-acoustic (LA) and transverse-acoustic (TA) branches are indicated. The dispersion relations are exactly those of the model material system used in this work. Specific details of the material system can be found in Ref.~\cite{maranets2024influence}.}
    \label{fig:dispersion_diagram}
\end{figure}

An important factor in designing thermal devices utilizing phononic crystals is the device length, sometimes referred to as the system size. The role of device length on heat conduction in periodic SLs, structures with an ordered interface arrangement, has been investigated in previous literature. Extensive studies, both experimental and computational, show the thermal conductivity of periodic SLs possessing period widths of just a few atomic layers increases with increasing device length \cite{chen1998thermal,Chen_1999,simkin2000minimum,daly2002molecular,luckyanova2012coherent,wang2014decomposition,wang2015optimization,luckyanova2018phonon,yu2019investigation,chakraborty2020complex,felix2020suppression,liu2022heat}. This behavior is characteristic of the ballistic propagation of coherent phonons (the resonant modes belonging to the dispersion relation of the phononic crystal itself as illustrated in Figs.~\ref{fig:dispersion_diagram}b and~\ref{fig:dispersion_diagram}d) whose mean free paths are bounded by the device length \cite{yang2003partially,wang2014decomposition,latour2014microscopic,maranets2024influence}. In contrast, for larger-period periodic SLs where incoherent phonons (modes following the bulk dispersion relations of the constituent layer materials as illustrated in Figs.~\ref{fig:dispersion_diagram}a and~\ref{fig:dispersion_diagram}c) possessing mean free paths bounded by the layer widths dominate heat conduction, thermal conductivity is minimally affected by device length beyond a few periods due to diffuse transmission of phonons scattered at the interfaces \cite{yang2003partially,luckyanova2012coherent,wang2014decomposition,latour2014microscopic,ma2022ex,maranets2024prominent}. For aperiodic SLs, structures with a disordered interface arrangement, several studies show thermal conductivity possesses a weaker dependence on length in comparison to periodic SLs, suggesting weaker phonon coherence \cite{wang2014decomposition,wang2015optimization,qiu2015effects,chakraborty2017ultralow,chakraborty2020quenching,hu2021direct,chakraborty2020complex}.

Several critical questions remain unanswered following these previous works. Firstly, how does the impact of device length on heat conduction in aperiodic SLs differ from that in periodic SLs? Secondly, what are the underlying mechanisms driving these behaviors in both periodic and aperiodic SLs? The latter question is particularly significant, as most studies have analyzed phonon coherence in SLs and other phononic crystals indirectly, often through computational predictions or experimental measurements of the overall thermal conductivity. While spectral analyses using methods such as the Boltzmann transport equation, lattice dynamics calculations, atomistic Green’s function, and normal mode analysis from finite-temperature molecular dynamics simulations provide valuable insights into phonon physics, these approaches are limited in their ability to capture or resolve the wave nature of phonons and dynamic wave phenomena, such as the establishment of phonon interference \cite{luo2013nanoscale,minnich2015advances,lindsay2019perspective}. 

To overcome these limitations, we employ the atomistic phonon wave-packet method, which is uniquely suitable for studying phonon coherence due to its capability to directly model the dynamic wave interference of phonons \cite{schelling2002phonon,schelling2003multiscale,latour2017distinguishing}. Using this method, we recently demonstrated that the previously observed reduction in the thermal conductivity of aperiodic SLs with decreasing temperature can be attributed to the diminished transmission of coherent phonons when their spatial coherence length increases (as coherence length is inversely related to temperature) \cite{latour2014microscopic,latour2017distinguishing,maranets2024influence}. Additionally, in a related wave-packet study, we provided direct evidence of the mode-conversion of incoherent phonons into coherent phonons governed by the dispersion relation of periodic SLs\textemdash a widely accepted hypothesis that had not been directly proven \cite{maranets2024prominent}. Furthermore, we revealed that this coherent mode-conversion effect also occurs in aperiodic SLs, promoting high phonon transmission and contributing to significant thermal conductivity that was previously unexplained \cite{maranets2024prominent}. 

Other studies employing the wave-packet technique have similarly uncovered novel physical phenomena that clarify trends in the thermophysical properties of various nanostructured materials where phonon wave effects are prominent \cite{schelling2003multiscale,tian2010phonon,liang2017phonon,shao2018understanding,shi2018dominant,desmarchelier2021ballistic,jiang2021total,maranets2023ballistic,beardo2024resonant}. Building upon this foundation, the present work aims to elucidate how phonon coherence evolves with device length in both periodic and aperiodic SLs, particularly in the context of heat conduction.

\section{Methodology}

\begin{figure}
    \centering
    \includegraphics[width=0.8\textwidth]{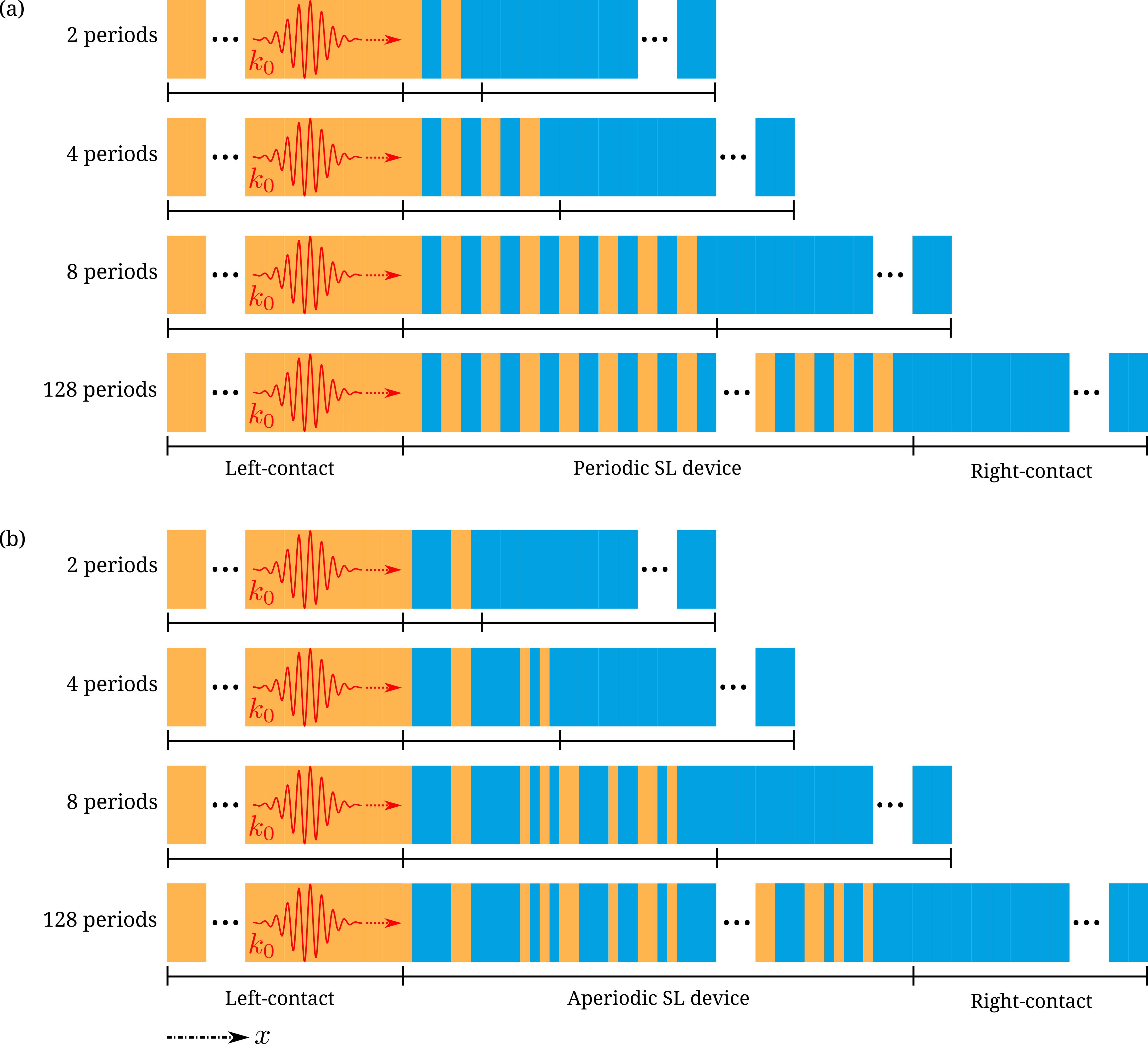}
    \caption{Schematic illustrations of the simulation domains for the 2 period, 4 period, 8 period, and 128 period device length cases for the periodic SL (a) and aperiodic SL (b); visualizing the wave-packet simulation process and the length reduction methodology. The incident LA-mode incoherent phonon wave-packet centered at wavevector $k_{0}$ in reciprocal-space is generated in the pure material left-contact and allowed to propagate into the SL device. The total energies of the left- and right-contacts, as well the device are recorded during the simulation to compute the transmission across the device. The sizes of the illustrated wavelength and wave-packet are not to scale.}
    \label{fig:system_length_diagram}
\end{figure}

The wave-packet technique applied in this study is adapted from the method by Schelling et al. \cite{schelling2002phonon} The system configuration is similar to our previous study in that we used the same model material system and simulate the scattering of an incoherent phonon belonging to the bulk dispersion relation of one of the constituent materials of the SL (Fig.~\ref{fig:dispersion_diagram}c) \cite{maranets2024prominent}. A phonon wave-packet is described by the following equation for the displacement of the $i$th atom in the $n$th unit cell.
\begin{equation}
u_{i,n} = \frac{A_{i}}{\sqrt{m_{i}}}\varepsilon_{k_{0},i}\exp{(i[k_{0}\cdot(x_{n} - x_{0})-\omega_{0}t]})\exp{(-4(x_{n}-x_{0}-v_{g0}t)^{2}/l_{c}^{2})}\qquad 
\label{eqn:pwp}
\end{equation}
$\varepsilon_{k_{0},i}$, $\omega_{0}$, and $v_{g0}$ are the eigenvector, frequency, and group velocity values associated with the phonon centered at wavevector $k_{0}$ in reciprocal-space. $m_{i}$ is the mass of atom $i$ and $x_{n}$ is the position of the $n$th unit cell. The wave amplitude $A_{i}$, the initial position of the wave-packet $x_{0}$, and the spatial coherence length $l_{c}$ are values specified by the researcher. The real parts of Eqn.~\ref{eqn:pwp} and the time-derivative of Eqn.~\ref{eqn:pwp} evaluated at time $t=0$ are applied as the initial atom displacement and velocity, respectively to generate the wave-packet. The simulations are conducted using the LAMMPS software \cite{thompson2022lammps}. Simulation parameters, such as the boundary conditions, simulation duration criterion, and transmission calculation methodology, are the same as Ref.~\cite{maranets2024prominent}.

To evaluate the role of device length on phonon coherence, we investigate how the transmission spectra of both periodic and aperiodic SL devices vary with device length. Specifically, we examine the effects of reducing the device length down to 1/64th of its initial size. The initial device length is set to 128 periods. Length reduction is implemented by sequentially shortening the structure by a specified fraction from the boundary at the right contact, as illustrated in Fig.~\ref{fig:system_length_diagram}. Using this approach, a device reduced to 1/2 of its initial length retains the first 64 periods starting from the left contact, while a device reduced to 1/64 of its initial length consists of only the first 2 periods.

For all simulations, the incident wave-packet is assigned a spatial coherence length four times the initial device length of 128 periods to ensure sufficient interference, as established in our prior study \cite{maranets2024influence}. Findings from that work further indicate that phonon transmission remains largely unaffected by the increasing disparity between the coherence length and the device length as the device is shortened \cite{maranets2024influence}. The interface arrangement in the aperiodic SL follows the methodology outlined in Ref.~\cite{wang2015optimization}, and all structures are assumed to have perfect interfaces without roughness.

\section{Results and Discussion}

\subsection{Periodic superlattice}

\begin{figure}
    \centering
    \includegraphics[width=0.9\textwidth]{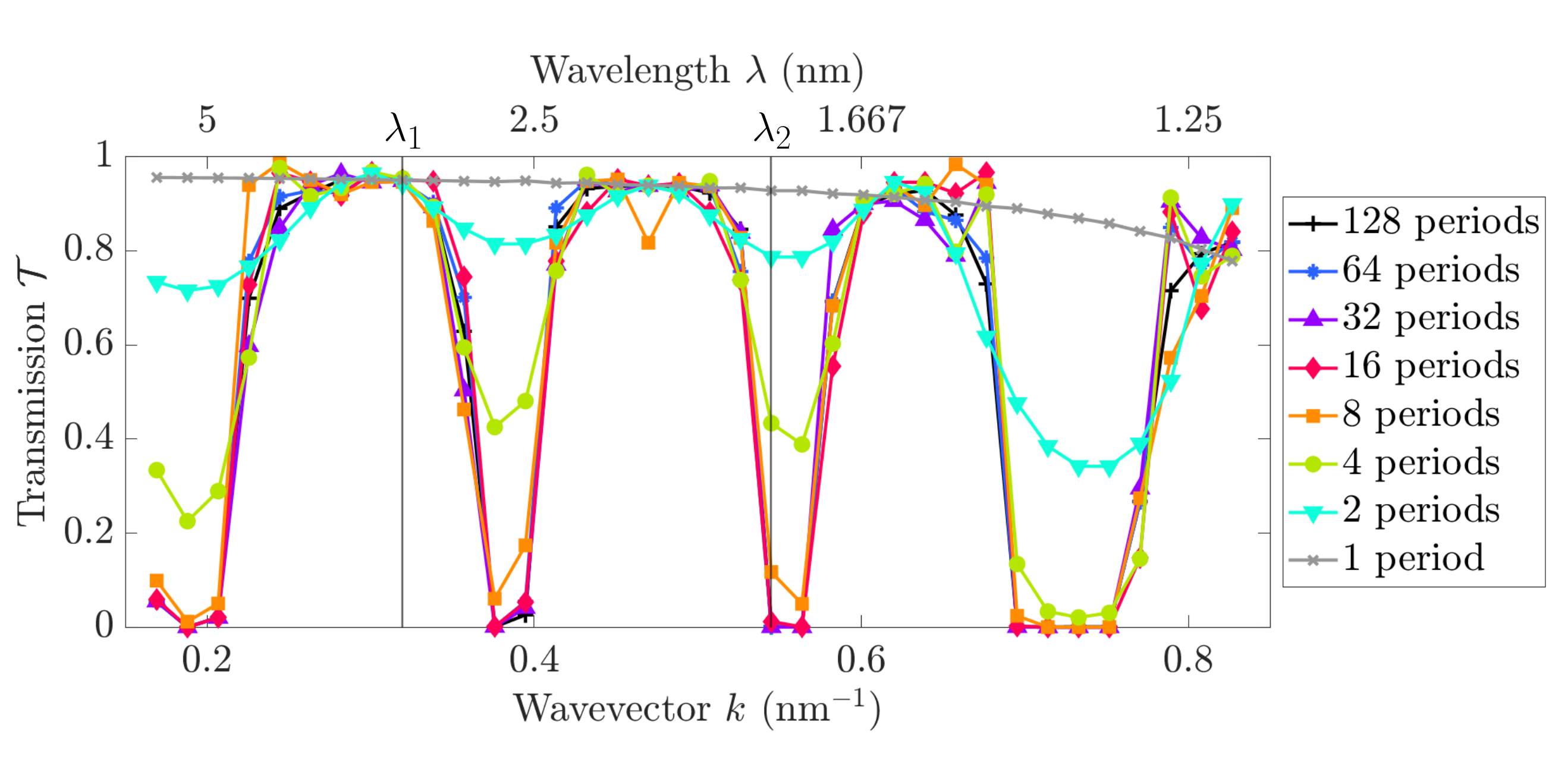}
    \caption{Transmission $\mathcal{T}$ versus wavevector $k$ (or inverse of wavelength $\lambda$) for the LA-mode incoherent phonon wave-packet propagating through periodic SL devices of varying length as shown in Fig.~\ref{fig:system_length_diagram}a. The 1 period case corresponds to transmission across a single interface as illustrated in Fig.~\ref{fig:dispersion_diagram}a. Transmission is computed per the approach in Ref.~\cite{maranets2024influence}. Key wavelengths $\lambda_{1} = 3.13$ nm and $\lambda_{2} = 1.83$ nm analyzed in this paper are indicated by the vertical black lines in the figure.}
    \label{fig:transmission_SL_length}
\end{figure}

We begin with the transmission through the periodic SL devices of varying length as shown in Fig.~\ref{fig:transmission_SL_length}. We find that the spectra nearly remains the same from the initial device length of 128 periods down to a 1/8th fraction of 16 periods. The regularly spaced zero transmission dips evidences the Bragg reflection of phonons previously reported in first-principles analyses and phonon imaging experiments as well as our previous studies \cite{tamura1988acoustic,tamura1988acousticmulti,tamura1989localized,hurley1987imaging,tanaka1998phonon,maranets2024prominent}. The high transmission values are a consequence of coherent mode-conversion where incoherent phonons transform through constructive interference to coherent phonons following the dispersion relation of the periodic SL. The lack of variance in the transmission spectra above 16 periods agrees with previous wave-packet studies of incoherent phonon transport in periodic SLs which showed unchanging transmission for select wavevectors above a certain device length \cite{schelling2003multiscale,jiang2021total}. Below 16 periods, we see the transmission dips dissipate with decreasing device length while the high transmission values remain roughly the same. These results suggest that the Bragg scattering condition in the periodic SL requires some minimum distance to fully manifest. We note this result was only partially observed in the previous aforementioned wave-packet studies which did not examine the transmission length-dependence for the entire phonon spectrum.

To illuminate the mechanisms of transmission length-dependence in the periodic SL, we analyze the phonon behaviors in reciprocal-space with the wavelet transform. The wavelet transform allows us to study the changes in the phonon wave-packet in reciprocal-space as a function of real-space position \cite{baker2012application}. Consequentially, this method is suitable for identifying mode-conversion considering we know the central wavevector $k_{0}$ of the incident phonon. 

\begin{figure}
    \centering
    \includegraphics[width=\textwidth]{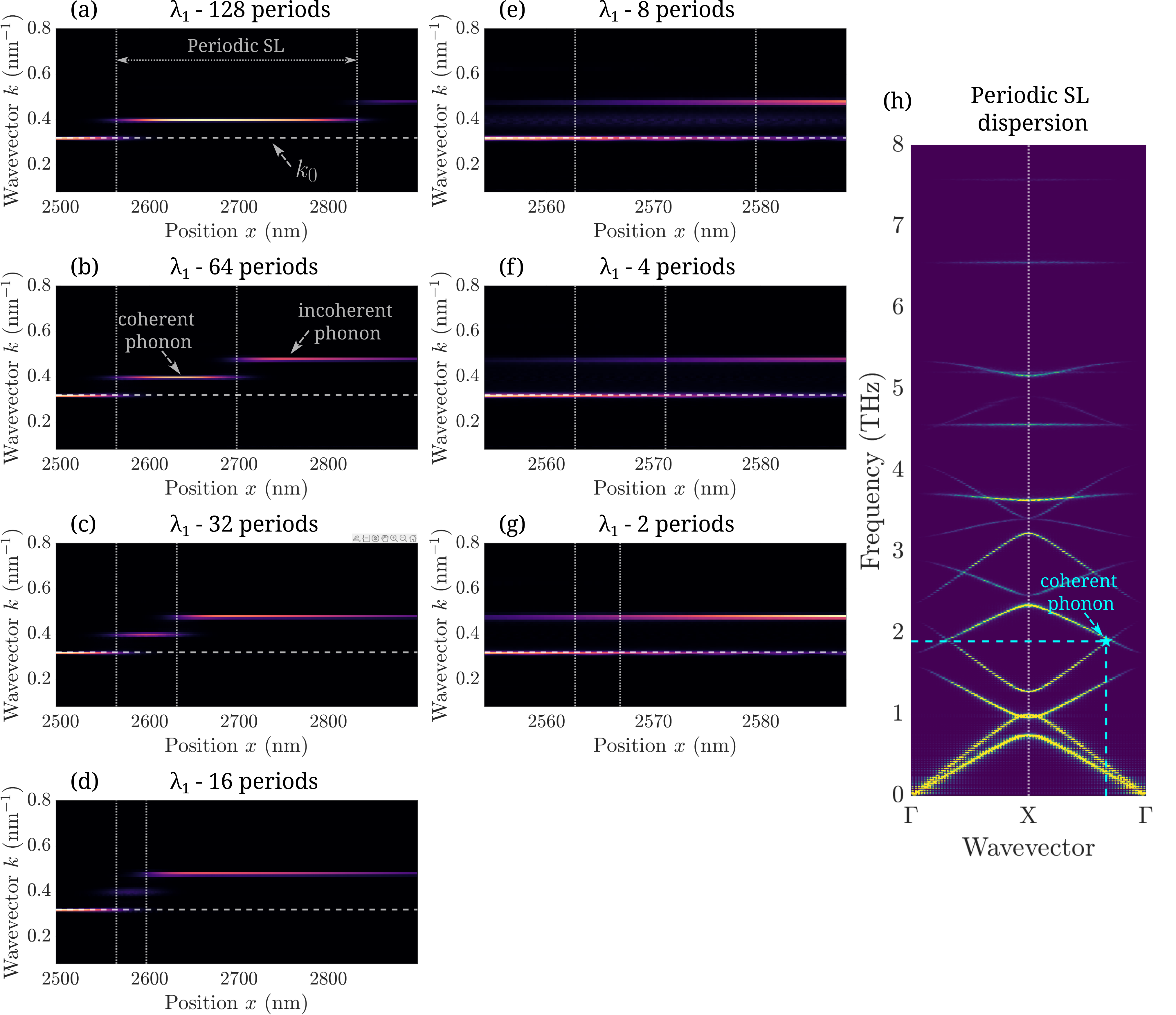}
    \caption{(a-g): Snapshots of the reciprocal-space wavelet transform when the incoherent phonon wave-packet scatters through the periodic SL device for an incident wavelength $\lambda_{1}=3.13$ nm. The dotted vertical lines denote the position and length of the periodic SL device. The dashed horizontal line denotes the central wavevector $k_{0}$ of the incident phonon. Illuminated regions in the heat map outside the $k_{0}$ line indicate mode-conversion. (h): Dispersion relation of the periodic SL computed from spectral energy density calculations with the mode-converted phonon in the $\lambda_{1}=3.13$ nm case located with the cyan marker and lines \cite{thomas2010predicting}.}
    \label{fig:wavelet_lambda1}
\end{figure}

For the periodic SL, we specifically examine the transport of two wavelengths indicated in Fig.~\ref{fig:transmission_SL_length}: $\lambda_{1}=3.13$ nm and $\lambda_{2}=1.83$ nm. These two wavelengths correspond to instances of near-unity transmission for all device lengths and dissipation of the zero-transmission Bragg reflection as device length is reduced, respectively. Plots of the wavelet transform for $\lambda_{1}$ for all device lengths are presented in Fig.~\ref{fig:wavelet_lambda1}. We provide the plots for $\lambda_{2}$ in Fig. S1.

For $\lambda_{1}$, at device lengths of 128 periods to 16 periods (Figs.~\ref{fig:wavelet_lambda1}a-\ref{fig:wavelet_lambda1}d), we find high transmission is facilitated by distinct coherent mode-conversion as observed in our previous study \cite{maranets2024prominent}. Fig.~\ref{fig:wavelet_lambda1}h provides the dispersion relation of the periodic SL with the coherent phonon indicated. For device lengths lower than 16 periods (Figs.~\ref{fig:wavelet_lambda1}e-\ref{fig:wavelet_lambda1}g), the coherent phonon gradually disappears and quick conversion to the corresponding incoherent phonon in the right-contact becomes more prominent. This result evidences the transition between coherent and incoherent phonon dominated transport at short device lengths. When there are few interfaces, the device is not long enough to support sufficient constructive interference to the resonant modes of the periodic SL, however, transmission is not attenuated as there now is less impedance for the particle-like transport of incoherent phonons. 

For $\lambda_{2}$, at device lengths of 128 periods to 16 periods (Figs. S1a-S1d), we find the zero-transmission is due to an absence of mode-conversion inside the periodic SL. As $\lambda_{2}$ satisfies the Bragg scattering condition, significant destructive interference of phonon waves results in total suppression  of energy transport. Below 16 periods (Figs. S1e-S1g), we observe increasingly prominent conversion to the corresponding incoherent phonon in the right-contact as device length is reduced, evidencing particle-like incoherent phonon transport similar to the $\lambda_{1}$ case. Most interestingly, the transmission of the $\lambda_{2}$ phonon is clearly less than the $\lambda_{1}$ phonon at very short device lengths, despite having very comparable transmission values in the single interface (i.e. 1 period) scenario. This result suggests that while phonon coherence (including the Bragg scattering effect) requires some distance to fully develop, it still affects energy transport even at the lowest device length of 2 periods. 

\subsection{Aperiodic superlattice}

\begin{figure}
    \centering
    \includegraphics[width=0.9\textwidth]{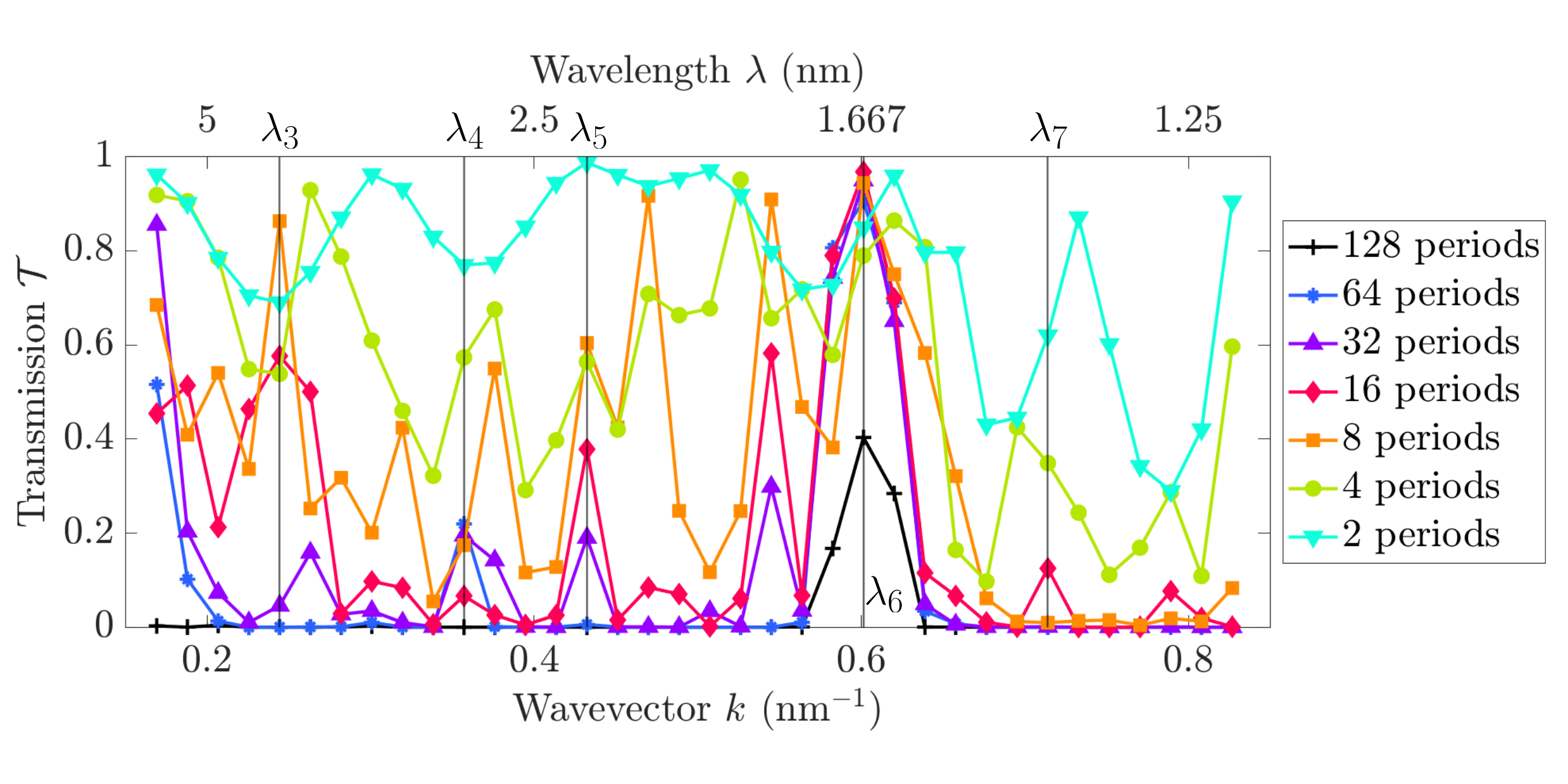}
    \caption{Transmission $\mathcal{T}$ versus wavevector $k$ (or inverse of wavelength $\lambda$) for the LA-mode incoherent phonon wave-packet propagating through aperiodic SL devices of varying length as shown in Fig.~\ref{fig:system_length_diagram}b. Transmission is computed per the approach in Ref.~\cite{maranets2024influence}. Key wavelengths $\lambda_{3}=4.09$ nm, $\lambda_{4}=2.80$ nm, $\lambda_{5}=2.31$ nm, $\lambda_{6}=1.66$ nm and $\lambda_{7}=1.40$ nm analyzed in this paper are indicated by the vertical black lines in the figure.}
    \label{fig:transmission_RML_length}
\end{figure}

We now turn our attention to the aperiodic SL. Unlike the periodic SL, Fig.~\ref{fig:transmission_RML_length} shows that the transmission spectra of aperiodic SL devices with reduced length deviates from the pattern of the 128 period device much earlier than 16 periods. As reported in our previous study, the prominent transmission peak at 1.66 nm is a consequence of mode-conversion to a vibrational mode belonging to the dispersion relation of the aperiodic SL, a behavior extensively observed in the periodic SL \cite{maranets2024prominent}. The zero-transmission values at other wavelengths for the long device lengths are indicative of an inability to mode-convert and the significant suppression of particle-like phonon transport due to the many interfaces in the device. The higher transmission values across the spectrum as the device length decreases suggest that either of these two effects or both, weakens with the reduction in device length. 

\begin{figure}
    \centering
    \includegraphics[width=\textwidth]{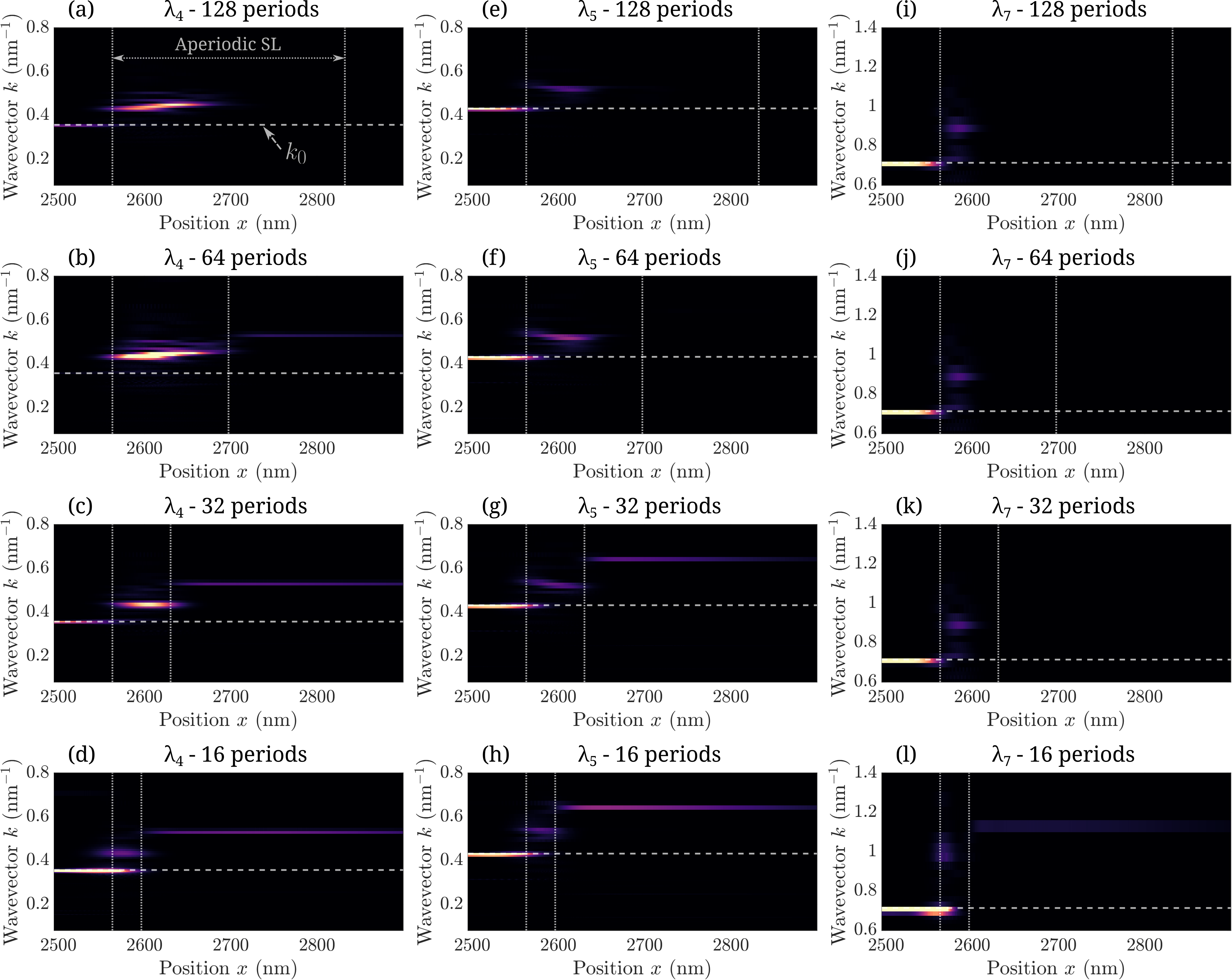}
    \caption{(a-l): Snapshots of the reciprocal-space wavelet transform when the incoherent phonon wave-packet scatters through the aperiodic SL device (128 to 16 periods) for incident wavelengths $\lambda_{4}=2.80$ nm, $\lambda_{5}=2.31$ nm, and $\lambda_{7}=1.40$ nm. The dotted vertical lines denote the position and length of the aperiodic SL device. The dashed horizontal line denotes the central wavevector $k_{0}$ of the incident phonon. Illuminated regions in the heat map outside the $k_{0}$ line indicate mode-conversion.}
    \label{fig:wavelet_RML}
\end{figure}

Similar to our analysis of the periodic SL, we compute the wavelet transforms for several wavelengths of interest marked in Fig.~\ref{fig:transmission_RML_length}. We are primarily interested in $\lambda_{4}=2.80$ nm, $\lambda_{5}=2.31$ nm, and $\lambda_{7}=1.40$ nm as these are cases of distinct transmission peaks, likely due to coherent mode-conversion, emerging as device length is reduced from the full length of 128 periods. Wavelet transforms of these wavelengths for device lengths of 128 to 16 periods are presented in Fig.~\ref{fig:wavelet_RML}. Remaining device lengths (8 to 2 periods) are provided in Fig. S2.

For 128 periods, $\lambda_{6}=1.66$ nm (wavelet transforms provided in Figs. S3) is the only wavelength that transmits significantly through the device. When the device length is halved to 64 periods, a second transmission peak arises at $\lambda_{4}$. Examining the wavelet transforms for $\lambda_{4}$ (Figs.~\ref{fig:wavelet_RML}a-\ref{fig:wavelet_RML}d), we observe coherent mode-conversion for device lengths of 128 periods to 16 periods. However, the mode-converted phonon has a finite spatial extension i.e. coherence length that dictates the transmission through the aperiodic SL device. Specifically, the coherence length is less than 128 periods but greater than 64 periods, enabling high transmission across the 64 period and smaller devices but not the 128 period device. This effect is similarly observed for wavelengths $\lambda_{5}$ (Figs.~\ref{fig:wavelet_RML}e-\ref{fig:wavelet_RML}h) and $\lambda_{7}$ (Figs.~\ref{fig:wavelet_RML}i-\ref{fig:wavelet_RML}l) that see additional transmission peaks emerge at 32 and 16 periods, respectively. In those cases, the coherent phonon inside the aperiodic SL has a coherence length greater than the device length at which the peak arises at but less than the larger device length scenarios.

The manifestation of coherent phonons that cannot span the entire device length is a behavior not observed in the periodic SL. The wavelet transforms for $\lambda_{1}$ (Fig.~\ref{fig:wavelet_lambda1}) show the mode-converted phonon always extends across the periodic SL device. Furthermore, comparing to $\lambda_{4}$, $\lambda_{5}$, and $\lambda_{7}$ visualized in Fig.~\ref{fig:wavelet_RML}, we see that the coherent phonon in the periodic SL is much narrower in reciprocal-space than the modes within aperiodic SL which appear more broadened. These results offer a critical clue as to why the length-dependence of transmission facilitated by coherent mode-conversion differs between the periodic and aperiodic SL. Firstly, the finite spatial extension of a phonon wave-packet and the broadening of its wavevector are interrelated \cite{latour2017distinguishing,maranets2024influence}. Secondly, the relation of spatial width to the characteristic length of a material, phononic or otherwise, determines the wave interference states and consequently, the phonon transport regime \cite{latour2014microscopic,latour2017distinguishing}. These facts, in conjunction with a new perspective on aperiodic SLs supported by our findings, can more rigorously explain the existing trends of thermal conductivity and other macroscopic properties that were previously just attributed to ``weakened phonon coherence" with minimal further explanation.

The aperiodic SL has been commonly described as the periodic SL with a perturbation to the layer thicknesses with the concept first being introduced by Tamura and Nori \cite{tamura1990acoustic}. As a result, heat conduction can be viewed as partially coherent since the coherent phonons belonging to the periodic SL dispersion relation now experience some scattering throughout the structure due to the perturbation. We argue that the aperiodic SL should rather be seen as the disordered version of the periodic SL in the same way amorphous solids are described as disordered forms of perfect crystals. The principles of phonon transport in amorphous solids align with the phonon behaviors observed in aperiodic SLs in this work; particularly the connection between phonon character and coherence length. In amorphous solids, the disordered arrangement of atoms destroys the long-range lattice periodicity of the perfect crystal. Consequently, Fourier decomposition of atomic vibrations to plane wave solutions becomes ill-defined. Heat is no longer carried by ballistically propagating wave-packets, rather the phase correlations of non-propagating phonon modes in real-space results in a diffuse transmission of energy \cite{allen1993thermal,feldman1993thermal,allen1999diffusons,lv2016examining,seyf2017rethinking,simoncelli2019unified,isaeva2019modeling}. How the measure of the spatial correlations, i.e. spatial coherence length, relates to the length-scale of local disorder determines the magnitude of heat conduction \cite{bodapati2006vibrations,zhang2022coherence}. 

This framework of phonon transport can be analogized to the aperiodic SL. Akin to amorphous solids, the aperiodic SL is characterized as a disordered arrangement of interfaces breaking the long-range order of regularly spaced interfaces in the periodic SL. The notion of weakened phonon coherence in reality stems from coherent phonons within the aperiodic SL not being well-defined plane waves that propagate ballistically as in the periodic SL. The ballistic phonons within the periodic system will be spatially bounded by the total length of the device whereas the non-propagative phonons in the aperiodic structure can interfere with the local disorder and thus possess spatial extensions less than the system size. In the context of wavelet transforms from wave-packet simulations, these phonon behaviors would be represented as vibrational energy spanning the entire length of the periodic SL while being considerably narrow in reciprocal-space and vibrational energy with broadened wavevector having finite spatial extension less than the aperiodic SL device length. These results are clearly observed in Fig.~\ref{fig:wavelet_lambda1} and Fig.~\ref{fig:wavelet_RML}. 

The analogy between aperiodic SLs and amorphous solids is further supported by existing thermal conductivity data. Beyond a system size of a few periods (where incoherent phonons dominate), the rate of increase in thermal conductivity with device length is generally much greater in periodic SLs than in aperiodic SLs \cite{wang2014decomposition,wang2015optimization,qiu2015effects,hu2021direct,chakraborty2020complex}. This behavior substantiates the notion that, in aperiodic SLs, coherent phonons do not propagate as ballistic wave-packets whose heat-carrying capacity is limited solely by the system size, as is typically observed in periodic SLs and perfect crystals. Instead, coherent phonons in aperiodic SLs exist as non-propagative vibrations, interacting with one another and coupling with local disorder to produce diffuse energy transmission.

The significant variation in transmission with device length, as presented in Fig.~\ref{fig:transmission_RML_length}---marked by the emergence of a second transmission peak at $\lambda_{4}$ for the 64-period device---further supports this interpretation. This variation reflects the changing length scale of local disorder as the device length decreases, as illustrated in Fig.~\ref{fig:system_length_diagram}b. Before proceeding, we highlight two critical points to address potential areas of confusion. 

First, the diffuse nature of coherent phonons in aperiodic SLs is fundamentally different from the diffuse transport of incoherent phonons that scatter at individual interfaces. Importantly, we do not conflate coherent and incoherent phonons within aperiodic SLs. If these concepts were identical, we would not observe mode-conversion within the aperiodic SL devices. The wavelet transforms presented in this study unequivocally demonstrate the opposite. 

Second, while the perturbative perspective of aperiodic SLs is not entirely incorrect, it requires nuance. For instance, the first-principles analysis of phonon transmission in aperiodic SLs by Tamura and Nori aligns with some findings in this work, particularly their observation of zero transmission for long device lengths unless incoherent phonons undergo mode-conversion to coherent phonons \cite{tamura1990acoustic}. Additionally, the dispersion relation of aperiodic SLs obtained through spectral energy density calculations in our prior work relies on perturbative treatment \cite{maranets2024prominent}.

Lastly, we address the seemingly erratic transmission spectra of the aperiodic SL with short device lengths (8 periods and fewer). Specifically, transmission through these devices does not always increase with decreasing device length as is expected for incoherent phonon dominated-transport. The wavelet transforms for one such case, $\lambda_{3}=4.09$ nm (Fig. S4), shows no unique effects. We believe the contrasting transmission behavior is due to minor wave interference, both constructive and destructive, occurring. The atomistic wave-packet method, being inherently motivated to simulate phonon wave dynamics, does not model purely particle-like phonon transport as well as other methods such as Boltzmann transport equation.

\section{Conclusions}

Thermal conductivity in aperiodic SLs exhibits a much weaker dependence on device length compared to periodic SLs. This phenomenon has traditionally been attributed to weakened phonon coherence, where coherent phonons are scattered by the irregularly spaced interfaces. In this work, we propose a new perspective: the observed difference in length dependence arises from a fundamental difference in the nature of coherent phonon transport between the two structures. 

Using rigorous atomistic wave-packet simulations, we demonstrate that coherent phonons in aperiodic SLs exhibit spatial extensions limited to a specific number of SL layers. If the spatial extension of these phonons is smaller than the device length, their transmission is effectively prohibited. In contrast, periodic SLs support coherent phonons generated via mode-conversion that extend across all SL layers, allowing for ballistic transport through the entire structure and resulting in high transmission. 

We further contextualize these results by drawing an analogy between periodic SLs and perfect crystals, and between aperiodic SLs and amorphous solids. In the aperiodic SLs, the disordered arrangement of interfaces—akin to atomic disorder in amorphous solids—renders the description of coherent phonons as ballistically propagating plane waves ill-defined. Instead, coherent phonons exist as non-propagative vibrational modes, interacting with local disorder and each other to enable diffuse energy transmission. 

The distinction between ballistic (periodic SL) and diffuse (aperiodic SL) phonon transport is reflected in the length dependence of phonon transmission computed in this study. These transmission results align with thermal conductivity trends reported in the literature. Our findings offer novel insights into the nature of phonon coherence and its implications for heat conduction in superlattices, with potential significance for the thermal design of nanostructures incorporating SLs.

\begin{acknowledgments}
The authors gratefully acknowledge the financial support from the National Science Foundation (CBET-2047109). Maranets thanks the support from the Nevada NASA Space Grant Graduate Research Opportunity Fellowship and the Nuclear Power Graduate Fellowship from the Nuclear Regulatory Commission. Additionally, the authors would like to acknowledge the support provided by the Research and Innovation team and the Cyberinfrastructure Team in the Office of Information Technology at the University of Nevada, Reno, for facilitating access to the Pronghorn High-Performance Computing Cluster.
\end{acknowledgments}

\section*{Supplementary Material}
Supplementary Material contains results that are not included in the main text but may help the readers understand the ideas presented in this manuscript. 

\section*{Author Contributions}
\textbf{Theodore Maranets:} Writing – review \& editing (equal), Writing – original draft (lead), Software (lead), Methodology (lead), Formal analysis (equal), Conceptualization (equal). \textbf{Yan Wang:} Writing – review \& editing (equal), Writing – original draft (equal), Supervision (lead), Funding acquisition (lead), Formal analysis (equal), Conceptualization (equal).

\section*{Data Availability}
The data that support the findings of this study are available from the corresponding author upon reasonable request.

%\nocite{*}
\bibliography{references}% Produces the bibliography via BibTeX.

\begin{thebibliography}{68}%
\makeatletter
\providecommand \@ifxundefined [1]{%
 \@ifx{#1\undefined}
}%
\providecommand \@ifnum [1]{%
 \ifnum #1\expandafter \@firstoftwo
 \else \expandafter \@secondoftwo
 \fi
}%
\providecommand \@ifx [1]{%
 \ifx #1\expandafter \@firstoftwo
 \else \expandafter \@secondoftwo
 \fi
}%
\providecommand \natexlab [1]{#1}%
\providecommand \enquote  [1]{``#1''}%
\providecommand \bibnamefont  [1]{#1}%
\providecommand \bibfnamefont [1]{#1}%
\providecommand \citenamefont [1]{#1}%
\providecommand \href@noop [0]{\@secondoftwo}%
\providecommand \href [0]{\begingroup \@sanitize@url \@href}%
\providecommand \@href[1]{\@@startlink{#1}\@@href}%
\providecommand \@@href[1]{\endgroup#1\@@endlink}%
\providecommand \@sanitize@url [0]{\catcode `\\12\catcode `\$12\catcode `\&12\catcode `\#12\catcode `\^12\catcode `\_12\catcode `\%12\relax}%
\providecommand \@@startlink[1]{}%
\providecommand \@@endlink[0]{}%
\providecommand \url  [0]{\begingroup\@sanitize@url \@url }%
\providecommand \@url [1]{\endgroup\@href {#1}{\urlprefix }}%
\providecommand \urlprefix  [0]{URL }%
\providecommand \Eprint [0]{\href }%
\providecommand \doibase [0]{https://doi.org/}%
\providecommand \selectlanguage [0]{\@gobble}%
\providecommand \bibinfo  [0]{\@secondoftwo}%
\providecommand \bibfield  [0]{\@secondoftwo}%
\providecommand \translation [1]{[#1]}%
\providecommand \BibitemOpen [0]{}%
\providecommand \bibitemStop [0]{}%
\providecommand \bibitemNoStop [0]{.\EOS\space}%
\providecommand \EOS [0]{\spacefactor3000\relax}%
\providecommand \BibitemShut  [1]{\csname bibitem#1\endcsname}%
\let\auto@bib@innerbib\@empty
%</preamble>
\bibitem [{\citenamefont {Hussein}\ \emph {et~al.}(2014)\citenamefont {Hussein}, \citenamefont {Leamy},\ and\ \citenamefont {Ruzzene}}]{hussein2014dynamics}%
  \BibitemOpen
  \bibfield  {author} {\bibinfo {author} {\bibfnamefont {M.~I.}\ \bibnamefont {Hussein}}, \bibinfo {author} {\bibfnamefont {M.~J.}\ \bibnamefont {Leamy}},\ and\ \bibinfo {author} {\bibfnamefont {M.}~\bibnamefont {Ruzzene}},\ }\bibfield  {title} {\bibinfo {title} {Dynamics of phononic materials and structures: Historical origins, recent progress, and future outlook},\ }\href@noop {} {\bibfield  {journal} {\bibinfo  {journal} {Applied Mechanics Reviews}\ }\textbf {\bibinfo {volume} {66}},\ \bibinfo {pages} {040802} (\bibinfo {year} {2014})}\BibitemShut {NoStop}%
\bibitem [{\citenamefont {Maldovan}(2015)}]{maldovan2015phonon}%
  \BibitemOpen
  \bibfield  {author} {\bibinfo {author} {\bibfnamefont {M.}~\bibnamefont {Maldovan}},\ }\bibfield  {title} {\bibinfo {title} {Phonon wave interference and thermal bandgap materials},\ }\href@noop {} {\bibfield  {journal} {\bibinfo  {journal} {Nature materials}\ }\textbf {\bibinfo {volume} {14}},\ \bibinfo {pages} {667} (\bibinfo {year} {2015})}\BibitemShut {NoStop}%
\bibitem [{\citenamefont {Volz}\ \emph {et~al.}(2016)\citenamefont {Volz}, \citenamefont {Ordonez-Miranda}, \citenamefont {Shchepetov}, \citenamefont {Prunnila}, \citenamefont {Ahopelto}, \citenamefont {Pezeril}, \citenamefont {Vaudel}, \citenamefont {Gusev}, \citenamefont {Ruello}, \citenamefont {Weig} \emph {et~al.}}]{volz2016nanophononics}%
  \BibitemOpen
  \bibfield  {author} {\bibinfo {author} {\bibfnamefont {S.}~\bibnamefont {Volz}}, \bibinfo {author} {\bibfnamefont {J.}~\bibnamefont {Ordonez-Miranda}}, \bibinfo {author} {\bibfnamefont {A.}~\bibnamefont {Shchepetov}}, \bibinfo {author} {\bibfnamefont {M.}~\bibnamefont {Prunnila}}, \bibinfo {author} {\bibfnamefont {J.}~\bibnamefont {Ahopelto}}, \bibinfo {author} {\bibfnamefont {T.}~\bibnamefont {Pezeril}}, \bibinfo {author} {\bibfnamefont {G.}~\bibnamefont {Vaudel}}, \bibinfo {author} {\bibfnamefont {V.}~\bibnamefont {Gusev}}, \bibinfo {author} {\bibfnamefont {P.}~\bibnamefont {Ruello}}, \bibinfo {author} {\bibfnamefont {E.~M.}\ \bibnamefont {Weig}}, \emph {et~al.},\ }\bibfield  {title} {\bibinfo {title} {Nanophononics: state of the art and perspectives},\ }\href@noop {} {\bibfield  {journal} {\bibinfo  {journal} {The European Physical Journal B}\ }\textbf {\bibinfo {volume} {89}},\ \bibinfo {pages} {1} (\bibinfo {year} {2016})}\BibitemShut {NoStop}%
\bibitem [{\citenamefont {Zhang}\ \emph {et~al.}(2021)\citenamefont {Zhang}, \citenamefont {Guo}, \citenamefont {Bescond}, \citenamefont {Chen}, \citenamefont {Nomura},\ and\ \citenamefont {Volz}}]{zhang2021coherent}%
  \BibitemOpen
  \bibfield  {author} {\bibinfo {author} {\bibfnamefont {Z.}~\bibnamefont {Zhang}}, \bibinfo {author} {\bibfnamefont {Y.}~\bibnamefont {Guo}}, \bibinfo {author} {\bibfnamefont {M.}~\bibnamefont {Bescond}}, \bibinfo {author} {\bibfnamefont {J.}~\bibnamefont {Chen}}, \bibinfo {author} {\bibfnamefont {M.}~\bibnamefont {Nomura}},\ and\ \bibinfo {author} {\bibfnamefont {S.}~\bibnamefont {Volz}},\ }\bibfield  {title} {\bibinfo {title} {Coherent thermal transport in nano-phononic crystals: An overview},\ }\href@noop {} {\bibfield  {journal} {\bibinfo  {journal} {APL Materials}\ }\textbf {\bibinfo {volume} {9}} (\bibinfo {year} {2021})}\BibitemShut {NoStop}%
\bibitem [{\citenamefont {Anufriev}\ \emph {et~al.}(2021)\citenamefont {Anufriev}, \citenamefont {Maire},\ and\ \citenamefont {Nomura}}]{anufriev2021review}%
  \BibitemOpen
  \bibfield  {author} {\bibinfo {author} {\bibfnamefont {R.}~\bibnamefont {Anufriev}}, \bibinfo {author} {\bibfnamefont {J.}~\bibnamefont {Maire}},\ and\ \bibinfo {author} {\bibfnamefont {M.}~\bibnamefont {Nomura}},\ }\bibfield  {title} {\bibinfo {title} {Review of coherent phonon and heat transport control in one-dimensional phononic crystals at nanoscale},\ }\href@noop {} {\bibfield  {journal} {\bibinfo  {journal} {APL Materials}\ }\textbf {\bibinfo {volume} {9}} (\bibinfo {year} {2021})}\BibitemShut {NoStop}%
\bibitem [{\citenamefont {Ma}(2023)}]{ma2023phonon}%
  \BibitemOpen
  \bibfield  {author} {\bibinfo {author} {\bibfnamefont {J.}~\bibnamefont {Ma}},\ }\bibfield  {title} {\bibinfo {title} {Phonon engineering of micro-and nanophononic crystals and acoustic metamaterials: A review},\ }\href@noop {} {\bibfield  {journal} {\bibinfo  {journal} {Small Science}\ }\textbf {\bibinfo {volume} {3}},\ \bibinfo {pages} {2200052} (\bibinfo {year} {2023})}\BibitemShut {NoStop}%
\bibitem [{\citenamefont {Chen}(1999)}]{Chen_1999}%
  \BibitemOpen
  \bibfield  {author} {\bibinfo {author} {\bibfnamefont {G.}~\bibnamefont {Chen}},\ }\bibfield  {title} {\bibinfo {title} {Phonon wave heat conduction in thin films and superlattices},\ }\href {https://doi.org/10.1115/1.2826085} {\bibfield  {journal} {\bibinfo  {journal} {Journal of Heat Transfer}\ }\textbf {\bibinfo {volume} {121}},\ \bibinfo {pages} {945–953} (\bibinfo {year} {1999})}\BibitemShut {NoStop}%
\bibitem [{\citenamefont {Simkin}\ and\ \citenamefont {Mahan}(2000)}]{simkin2000minimum}%
  \BibitemOpen
  \bibfield  {author} {\bibinfo {author} {\bibfnamefont {M.}~\bibnamefont {Simkin}}\ and\ \bibinfo {author} {\bibfnamefont {G.}~\bibnamefont {Mahan}},\ }\bibfield  {title} {\bibinfo {title} {Minimum thermal conductivity of superlattices},\ }\href@noop {} {\bibfield  {journal} {\bibinfo  {journal} {Physical Review Letters}\ }\textbf {\bibinfo {volume} {84}},\ \bibinfo {pages} {927} (\bibinfo {year} {2000})}\BibitemShut {NoStop}%
\bibitem [{\citenamefont {Ravichandran}\ \emph {et~al.}(2014)\citenamefont {Ravichandran}, \citenamefont {Yadav}, \citenamefont {Cheaito}, \citenamefont {Rossen}, \citenamefont {Soukiassian}, \citenamefont {Suresha}, \citenamefont {Duda}, \citenamefont {Foley}, \citenamefont {Lee}, \citenamefont {Zhu} \emph {et~al.}}]{ravichandran2014crossover}%
  \BibitemOpen
  \bibfield  {author} {\bibinfo {author} {\bibfnamefont {J.}~\bibnamefont {Ravichandran}}, \bibinfo {author} {\bibfnamefont {A.~K.}\ \bibnamefont {Yadav}}, \bibinfo {author} {\bibfnamefont {R.}~\bibnamefont {Cheaito}}, \bibinfo {author} {\bibfnamefont {P.~B.}\ \bibnamefont {Rossen}}, \bibinfo {author} {\bibfnamefont {A.}~\bibnamefont {Soukiassian}}, \bibinfo {author} {\bibfnamefont {S.}~\bibnamefont {Suresha}}, \bibinfo {author} {\bibfnamefont {J.~C.}\ \bibnamefont {Duda}}, \bibinfo {author} {\bibfnamefont {B.~M.}\ \bibnamefont {Foley}}, \bibinfo {author} {\bibfnamefont {C.-H.}\ \bibnamefont {Lee}}, \bibinfo {author} {\bibfnamefont {Y.}~\bibnamefont {Zhu}}, \emph {et~al.},\ }\bibfield  {title} {\bibinfo {title} {Crossover from incoherent to coherent phonon scattering in epitaxial oxide superlattices},\ }\href@noop {} {\bibfield  {journal} {\bibinfo  {journal} {Nature materials}\ }\textbf {\bibinfo {volume} {13}},\ \bibinfo {pages} {168} (\bibinfo {year} {2014})}\BibitemShut {NoStop}%
\bibitem [{\citenamefont {Latour}\ \emph {et~al.}(2014)\citenamefont {Latour}, \citenamefont {Volz},\ and\ \citenamefont {Chalopin}}]{latour2014microscopic}%
  \BibitemOpen
  \bibfield  {author} {\bibinfo {author} {\bibfnamefont {B.}~\bibnamefont {Latour}}, \bibinfo {author} {\bibfnamefont {S.}~\bibnamefont {Volz}},\ and\ \bibinfo {author} {\bibfnamefont {Y.}~\bibnamefont {Chalopin}},\ }\bibfield  {title} {\bibinfo {title} {Microscopic description of thermal-phonon coherence: From coherent transport to diffuse interface scattering in superlattices},\ }\href@noop {} {\bibfield  {journal} {\bibinfo  {journal} {Physical Review B}\ }\textbf {\bibinfo {volume} {90}},\ \bibinfo {pages} {014307} (\bibinfo {year} {2014})}\BibitemShut {NoStop}%
\bibitem [{\citenamefont {Zhu}\ and\ \citenamefont {Ertekin}(2014)}]{zhu2014phonon}%
  \BibitemOpen
  \bibfield  {author} {\bibinfo {author} {\bibfnamefont {T.}~\bibnamefont {Zhu}}\ and\ \bibinfo {author} {\bibfnamefont {E.}~\bibnamefont {Ertekin}},\ }\bibfield  {title} {\bibinfo {title} {Phonon transport on two-dimensional graphene/boron nitride superlattices},\ }\href@noop {} {\bibfield  {journal} {\bibinfo  {journal} {Physical Review B}\ }\textbf {\bibinfo {volume} {90}},\ \bibinfo {pages} {195209} (\bibinfo {year} {2014})}\BibitemShut {NoStop}%
\bibitem [{\citenamefont {Xie}\ \emph {et~al.}(2018)\citenamefont {Xie}, \citenamefont {Ding},\ and\ \citenamefont {Zhang}}]{xie2018phonon}%
  \BibitemOpen
  \bibfield  {author} {\bibinfo {author} {\bibfnamefont {G.}~\bibnamefont {Xie}}, \bibinfo {author} {\bibfnamefont {D.}~\bibnamefont {Ding}},\ and\ \bibinfo {author} {\bibfnamefont {G.}~\bibnamefont {Zhang}},\ }\bibfield  {title} {\bibinfo {title} {Phonon coherence and its effect on thermal conductivity of nanostructures},\ }\href@noop {} {\bibfield  {journal} {\bibinfo  {journal} {Advances in Physics: X}\ }\textbf {\bibinfo {volume} {3}},\ \bibinfo {pages} {1480417} (\bibinfo {year} {2018})}\BibitemShut {NoStop}%
\bibitem [{\citenamefont {Chen}(2021)}]{chen2021non}%
  \BibitemOpen
  \bibfield  {author} {\bibinfo {author} {\bibfnamefont {G.}~\bibnamefont {Chen}},\ }\bibfield  {title} {\bibinfo {title} {Non-fourier phonon heat conduction at the microscale and nanoscale},\ }\href@noop {} {\bibfield  {journal} {\bibinfo  {journal} {Nature Reviews Physics}\ }\textbf {\bibinfo {volume} {3}},\ \bibinfo {pages} {555} (\bibinfo {year} {2021})}\BibitemShut {NoStop}%
\bibitem [{\citenamefont {Liu}\ \emph {et~al.}(2022)\citenamefont {Liu}, \citenamefont {Guo}, \citenamefont {Khvesyuk}, \citenamefont {Barinov},\ and\ \citenamefont {Wang}}]{liu2022heat}%
  \BibitemOpen
  \bibfield  {author} {\bibinfo {author} {\bibfnamefont {B.}~\bibnamefont {Liu}}, \bibinfo {author} {\bibfnamefont {Y.}~\bibnamefont {Guo}}, \bibinfo {author} {\bibfnamefont {V.~I.}\ \bibnamefont {Khvesyuk}}, \bibinfo {author} {\bibfnamefont {A.~A.}\ \bibnamefont {Barinov}},\ and\ \bibinfo {author} {\bibfnamefont {M.}~\bibnamefont {Wang}},\ }\bibfield  {title} {\bibinfo {title} {Heat conduction of multilayer nanostructures with consideration of coherent and incoherent phonon transport},\ }\href@noop {} {\bibfield  {journal} {\bibinfo  {journal} {Nano Research}\ }\textbf {\bibinfo {volume} {15}},\ \bibinfo {pages} {9492} (\bibinfo {year} {2022})}\BibitemShut {NoStop}%
\bibitem [{\citenamefont {Zhao}\ \emph {et~al.}(2022)\citenamefont {Zhao}, \citenamefont {Zhang}, \citenamefont {Song}, \citenamefont {Du}, \citenamefont {Wu}, \citenamefont {Kang},\ and\ \citenamefont {Sun}}]{zhao2022incoherent}%
  \BibitemOpen
  \bibfield  {author} {\bibinfo {author} {\bibfnamefont {L.}~\bibnamefont {Zhao}}, \bibinfo {author} {\bibfnamefont {L.}~\bibnamefont {Zhang}}, \bibinfo {author} {\bibfnamefont {H.}~\bibnamefont {Song}}, \bibinfo {author} {\bibfnamefont {H.}~\bibnamefont {Du}}, \bibinfo {author} {\bibfnamefont {J.}~\bibnamefont {Wu}}, \bibinfo {author} {\bibfnamefont {F.}~\bibnamefont {Kang}},\ and\ \bibinfo {author} {\bibfnamefont {B.}~\bibnamefont {Sun}},\ }\bibfield  {title} {\bibinfo {title} {Incoherent phonon transport dominates heat conduction across van der waals superlattices},\ }\href@noop {} {\bibfield  {journal} {\bibinfo  {journal} {Applied Physics Letters}\ }\textbf {\bibinfo {volume} {121}} (\bibinfo {year} {2022})}\BibitemShut {NoStop}%
\bibitem [{\citenamefont {Capinski}\ \emph {et~al.}(1999)\citenamefont {Capinski}, \citenamefont {Maris}, \citenamefont {Ruf}, \citenamefont {Cardona}, \citenamefont {Ploog},\ and\ \citenamefont {Katzer}}]{capinski1999thermal}%
  \BibitemOpen
  \bibfield  {author} {\bibinfo {author} {\bibfnamefont {W.}~\bibnamefont {Capinski}}, \bibinfo {author} {\bibfnamefont {H.}~\bibnamefont {Maris}}, \bibinfo {author} {\bibfnamefont {T.}~\bibnamefont {Ruf}}, \bibinfo {author} {\bibfnamefont {M.}~\bibnamefont {Cardona}}, \bibinfo {author} {\bibfnamefont {K.}~\bibnamefont {Ploog}},\ and\ \bibinfo {author} {\bibfnamefont {D.}~\bibnamefont {Katzer}},\ }\bibfield  {title} {\bibinfo {title} {Thermal-conductivity measurements of gaas/alas superlattices using a picosecond optical pump-and-probe technique},\ }\href@noop {} {\bibfield  {journal} {\bibinfo  {journal} {Physical Review B}\ }\textbf {\bibinfo {volume} {59}},\ \bibinfo {pages} {8105} (\bibinfo {year} {1999})}\BibitemShut {NoStop}%
\bibitem [{\citenamefont {Venkatasubramanian}(2000)}]{venkatasubramanian2000lattice}%
  \BibitemOpen
  \bibfield  {author} {\bibinfo {author} {\bibfnamefont {R.}~\bibnamefont {Venkatasubramanian}},\ }\bibfield  {title} {\bibinfo {title} {Lattice thermal conductivity reduction and phonon localizationlike behavior in superlattice structures},\ }\href@noop {} {\bibfield  {journal} {\bibinfo  {journal} {Physical Review B}\ }\textbf {\bibinfo {volume} {61}},\ \bibinfo {pages} {3091} (\bibinfo {year} {2000})}\BibitemShut {NoStop}%
\bibitem [{\citenamefont {Daly}\ \emph {et~al.}(2002)\citenamefont {Daly}, \citenamefont {Maris}, \citenamefont {Imamura},\ and\ \citenamefont {Tamura}}]{daly2002molecular}%
  \BibitemOpen
  \bibfield  {author} {\bibinfo {author} {\bibfnamefont {B.~C.}\ \bibnamefont {Daly}}, \bibinfo {author} {\bibfnamefont {H.~J.}\ \bibnamefont {Maris}}, \bibinfo {author} {\bibfnamefont {K.}~\bibnamefont {Imamura}},\ and\ \bibinfo {author} {\bibfnamefont {S.}~\bibnamefont {Tamura}},\ }\bibfield  {title} {\bibinfo {title} {Molecular dynamics calculation of the thermal conductivity of superlattices},\ }\href@noop {} {\bibfield  {journal} {\bibinfo  {journal} {Physical review B}\ }\textbf {\bibinfo {volume} {66}},\ \bibinfo {pages} {024301} (\bibinfo {year} {2002})}\BibitemShut {NoStop}%
\bibitem [{\citenamefont {Latour}\ and\ \citenamefont {Chalopin}(2017)}]{latour2017distinguishing}%
  \BibitemOpen
  \bibfield  {author} {\bibinfo {author} {\bibfnamefont {B.}~\bibnamefont {Latour}}\ and\ \bibinfo {author} {\bibfnamefont {Y.}~\bibnamefont {Chalopin}},\ }\bibfield  {title} {\bibinfo {title} {Distinguishing between spatial coherence and temporal coherence of phonons},\ }\href@noop {} {\bibfield  {journal} {\bibinfo  {journal} {Physical Review B}\ }\textbf {\bibinfo {volume} {95}},\ \bibinfo {pages} {214310} (\bibinfo {year} {2017})}\BibitemShut {NoStop}%
\bibitem [{\citenamefont {Maranets}\ and\ \citenamefont {Wang}(2024)}]{maranets2024prominent}%
  \BibitemOpen
  \bibfield  {author} {\bibinfo {author} {\bibfnamefont {T.}~\bibnamefont {Maranets}}\ and\ \bibinfo {author} {\bibfnamefont {Y.}~\bibnamefont {Wang}},\ }\bibfield  {title} {\bibinfo {title} {{Prominent phonon transmission across aperiodic superlattice through coherent mode-conversion}},\ }\href {https://doi.org/10.1063/5.0220824} {\bibfield  {journal} {\bibinfo  {journal} {Applied Physics Letters}\ }\textbf {\bibinfo {volume} {125}},\ \bibinfo {pages} {042205} (\bibinfo {year} {2024})}\BibitemShut {NoStop}%
\bibitem [{\citenamefont {Wang}\ \emph {et~al.}(2014)\citenamefont {Wang}, \citenamefont {Huang},\ and\ \citenamefont {Ruan}}]{wang2014decomposition}%
  \BibitemOpen
  \bibfield  {author} {\bibinfo {author} {\bibfnamefont {Y.}~\bibnamefont {Wang}}, \bibinfo {author} {\bibfnamefont {H.}~\bibnamefont {Huang}},\ and\ \bibinfo {author} {\bibfnamefont {X.}~\bibnamefont {Ruan}},\ }\bibfield  {title} {\bibinfo {title} {Decomposition of coherent and incoherent phonon conduction in superlattices and random multilayers},\ }\href@noop {} {\bibfield  {journal} {\bibinfo  {journal} {Physical Review B}\ }\textbf {\bibinfo {volume} {90}},\ \bibinfo {pages} {165406} (\bibinfo {year} {2014})}\BibitemShut {NoStop}%
\bibitem [{\citenamefont {Wang}\ \emph {et~al.}(2015)\citenamefont {Wang}, \citenamefont {Gu},\ and\ \citenamefont {Ruan}}]{wang2015optimization}%
  \BibitemOpen
  \bibfield  {author} {\bibinfo {author} {\bibfnamefont {Y.}~\bibnamefont {Wang}}, \bibinfo {author} {\bibfnamefont {C.}~\bibnamefont {Gu}},\ and\ \bibinfo {author} {\bibfnamefont {X.}~\bibnamefont {Ruan}},\ }\bibfield  {title} {\bibinfo {title} {Optimization of the random multilayer structure to break the random-alloy limit of thermal conductivity},\ }\href@noop {} {\bibfield  {journal} {\bibinfo  {journal} {Applied Physics Letters}\ }\textbf {\bibinfo {volume} {106}} (\bibinfo {year} {2015})}\BibitemShut {NoStop}%
\bibitem [{\citenamefont {Juntunen}\ \emph {et~al.}(2019)\citenamefont {Juntunen}, \citenamefont {V{\"a}nsk{\"a}},\ and\ \citenamefont {Tittonen}}]{juntunen2019anderson}%
  \BibitemOpen
  \bibfield  {author} {\bibinfo {author} {\bibfnamefont {T.}~\bibnamefont {Juntunen}}, \bibinfo {author} {\bibfnamefont {O.}~\bibnamefont {V{\"a}nsk{\"a}}},\ and\ \bibinfo {author} {\bibfnamefont {I.}~\bibnamefont {Tittonen}},\ }\bibfield  {title} {\bibinfo {title} {Anderson localization quenches thermal transport in aperiodic superlattices},\ }\href@noop {} {\bibfield  {journal} {\bibinfo  {journal} {Physical review letters}\ }\textbf {\bibinfo {volume} {122}},\ \bibinfo {pages} {105901} (\bibinfo {year} {2019})}\BibitemShut {NoStop}%
\bibitem [{\citenamefont {Hu}\ and\ \citenamefont {Tian}(2021)}]{hu2021direct}%
  \BibitemOpen
  \bibfield  {author} {\bibinfo {author} {\bibfnamefont {R.}~\bibnamefont {Hu}}\ and\ \bibinfo {author} {\bibfnamefont {Z.}~\bibnamefont {Tian}},\ }\bibfield  {title} {\bibinfo {title} {Direct observation of phonon anderson localization in si/ge aperiodic superlattices},\ }\href@noop {} {\bibfield  {journal} {\bibinfo  {journal} {Physical Review B}\ }\textbf {\bibinfo {volume} {103}},\ \bibinfo {pages} {045304} (\bibinfo {year} {2021})}\BibitemShut {NoStop}%
\bibitem [{\citenamefont {Maranets}\ \emph {et~al.}(2024)\citenamefont {Maranets}, \citenamefont {Nasiri},\ and\ \citenamefont {Wang}}]{maranets2024influence}%
  \BibitemOpen
  \bibfield  {author} {\bibinfo {author} {\bibfnamefont {T.}~\bibnamefont {Maranets}}, \bibinfo {author} {\bibfnamefont {M.}~\bibnamefont {Nasiri}},\ and\ \bibinfo {author} {\bibfnamefont {Y.}~\bibnamefont {Wang}},\ }\bibfield  {title} {\bibinfo {title} {Influence of spatial coherence on phonon transmission across aperiodically arranged interfaces},\ }\href@noop {} {\bibfield  {journal} {\bibinfo  {journal} {Physics Letters A}\ }\textbf {\bibinfo {volume} {512}},\ \bibinfo {pages} {129572} (\bibinfo {year} {2024})}\BibitemShut {NoStop}%
\bibitem [{\citenamefont {Shi}\ \emph {et~al.}(2015)\citenamefont {Shi}, \citenamefont {Dames}, \citenamefont {Lukes}, \citenamefont {Reddy}, \citenamefont {Duda}, \citenamefont {Cahill}, \citenamefont {Lee}, \citenamefont {Marconnet}, \citenamefont {Goodson}, \citenamefont {Bahk} \emph {et~al.}}]{shi2015evaluating}%
  \BibitemOpen
  \bibfield  {author} {\bibinfo {author} {\bibfnamefont {L.}~\bibnamefont {Shi}}, \bibinfo {author} {\bibfnamefont {C.}~\bibnamefont {Dames}}, \bibinfo {author} {\bibfnamefont {J.~R.}\ \bibnamefont {Lukes}}, \bibinfo {author} {\bibfnamefont {P.}~\bibnamefont {Reddy}}, \bibinfo {author} {\bibfnamefont {J.}~\bibnamefont {Duda}}, \bibinfo {author} {\bibfnamefont {D.~G.}\ \bibnamefont {Cahill}}, \bibinfo {author} {\bibfnamefont {J.}~\bibnamefont {Lee}}, \bibinfo {author} {\bibfnamefont {A.}~\bibnamefont {Marconnet}}, \bibinfo {author} {\bibfnamefont {K.~E.}\ \bibnamefont {Goodson}}, \bibinfo {author} {\bibfnamefont {J.-H.}\ \bibnamefont {Bahk}}, \emph {et~al.},\ }\bibfield  {title} {\bibinfo {title} {Evaluating broader impacts of nanoscale thermal transport research},\ }\href@noop {} {\bibfield  {journal} {\bibinfo  {journal} {Nanoscale and Microscale Thermophysical Engineering}\ }\textbf {\bibinfo {volume} {19}},\ \bibinfo {pages} {127} (\bibinfo {year} {2015})}\BibitemShut {NoStop}%
\bibitem [{\citenamefont {Chen}(1998)}]{chen1998thermal}%
  \BibitemOpen
  \bibfield  {author} {\bibinfo {author} {\bibfnamefont {G.}~\bibnamefont {Chen}},\ }\bibfield  {title} {\bibinfo {title} {Thermal conductivity and ballistic-phonon transport in the cross-plane direction of superlattices},\ }\href@noop {} {\bibfield  {journal} {\bibinfo  {journal} {Physical Review B}\ }\textbf {\bibinfo {volume} {57}},\ \bibinfo {pages} {14958} (\bibinfo {year} {1998})}\BibitemShut {NoStop}%
\bibitem [{\citenamefont {Luckyanova}\ \emph {et~al.}(2012)\citenamefont {Luckyanova}, \citenamefont {Garg}, \citenamefont {Esfarjani}, \citenamefont {Jandl}, \citenamefont {Bulsara}, \citenamefont {Schmidt}, \citenamefont {Minnich}, \citenamefont {Chen}, \citenamefont {Dresselhaus}, \citenamefont {Ren} \emph {et~al.}}]{luckyanova2012coherent}%
  \BibitemOpen
  \bibfield  {author} {\bibinfo {author} {\bibfnamefont {M.~N.}\ \bibnamefont {Luckyanova}}, \bibinfo {author} {\bibfnamefont {J.}~\bibnamefont {Garg}}, \bibinfo {author} {\bibfnamefont {K.}~\bibnamefont {Esfarjani}}, \bibinfo {author} {\bibfnamefont {A.}~\bibnamefont {Jandl}}, \bibinfo {author} {\bibfnamefont {M.~T.}\ \bibnamefont {Bulsara}}, \bibinfo {author} {\bibfnamefont {A.~J.}\ \bibnamefont {Schmidt}}, \bibinfo {author} {\bibfnamefont {A.~J.}\ \bibnamefont {Minnich}}, \bibinfo {author} {\bibfnamefont {S.}~\bibnamefont {Chen}}, \bibinfo {author} {\bibfnamefont {M.~S.}\ \bibnamefont {Dresselhaus}}, \bibinfo {author} {\bibfnamefont {Z.}~\bibnamefont {Ren}}, \emph {et~al.},\ }\bibfield  {title} {\bibinfo {title} {Coherent phonon heat conduction in superlattices},\ }\href@noop {} {\bibfield  {journal} {\bibinfo  {journal} {Science}\ }\textbf {\bibinfo {volume} {338}},\ \bibinfo {pages} {936} (\bibinfo {year} {2012})}\BibitemShut {NoStop}%
\bibitem [{\citenamefont {Luckyanova}\ \emph {et~al.}(2018)\citenamefont {Luckyanova}, \citenamefont {Mendoza}, \citenamefont {Lu}, \citenamefont {Song}, \citenamefont {Huang}, \citenamefont {Zhou}, \citenamefont {Li}, \citenamefont {Dong}, \citenamefont {Zhou}, \citenamefont {Garlow} \emph {et~al.}}]{luckyanova2018phonon}%
  \BibitemOpen
  \bibfield  {author} {\bibinfo {author} {\bibfnamefont {M.~N.}\ \bibnamefont {Luckyanova}}, \bibinfo {author} {\bibfnamefont {J.}~\bibnamefont {Mendoza}}, \bibinfo {author} {\bibfnamefont {H.}~\bibnamefont {Lu}}, \bibinfo {author} {\bibfnamefont {B.}~\bibnamefont {Song}}, \bibinfo {author} {\bibfnamefont {S.}~\bibnamefont {Huang}}, \bibinfo {author} {\bibfnamefont {J.}~\bibnamefont {Zhou}}, \bibinfo {author} {\bibfnamefont {M.}~\bibnamefont {Li}}, \bibinfo {author} {\bibfnamefont {Y.}~\bibnamefont {Dong}}, \bibinfo {author} {\bibfnamefont {H.}~\bibnamefont {Zhou}}, \bibinfo {author} {\bibfnamefont {J.}~\bibnamefont {Garlow}}, \emph {et~al.},\ }\bibfield  {title} {\bibinfo {title} {Phonon localization in heat conduction},\ }\href@noop {} {\bibfield  {journal} {\bibinfo  {journal} {Science advances}\ }\textbf {\bibinfo {volume} {4}},\ \bibinfo {pages} {eaat9460} (\bibinfo {year} {2018})}\BibitemShut {NoStop}%
\bibitem [{\citenamefont {Yu}\ \emph {et~al.}(2019)\citenamefont {Yu}, \citenamefont {Li},\ and\ \citenamefont {Ye}}]{yu2019investigation}%
  \BibitemOpen
  \bibfield  {author} {\bibinfo {author} {\bibfnamefont {J.}~\bibnamefont {Yu}}, \bibinfo {author} {\bibfnamefont {Q.}~\bibnamefont {Li}},\ and\ \bibinfo {author} {\bibfnamefont {W.}~\bibnamefont {Ye}},\ }\bibfield  {title} {\bibinfo {title} {Investigation of wave interference effect in si/ge superlattices with interfering monte carlo method},\ }\href@noop {} {\bibfield  {journal} {\bibinfo  {journal} {International Journal of Heat and Mass Transfer}\ }\textbf {\bibinfo {volume} {128}},\ \bibinfo {pages} {270} (\bibinfo {year} {2019})}\BibitemShut {NoStop}%
\bibitem [{\citenamefont {Chakraborty}\ \emph {et~al.}(2020{\natexlab{a}})\citenamefont {Chakraborty}, \citenamefont {Chiu}, \citenamefont {Ma},\ and\ \citenamefont {Wang}}]{chakraborty2020complex}%
  \BibitemOpen
  \bibfield  {author} {\bibinfo {author} {\bibfnamefont {P.}~\bibnamefont {Chakraborty}}, \bibinfo {author} {\bibfnamefont {I.~A.}\ \bibnamefont {Chiu}}, \bibinfo {author} {\bibfnamefont {T.}~\bibnamefont {Ma}},\ and\ \bibinfo {author} {\bibfnamefont {Y.}~\bibnamefont {Wang}},\ }\bibfield  {title} {\bibinfo {title} {Complex temperature dependence of coherent and incoherent lattice thermal transport in superlattices},\ }\href@noop {} {\bibfield  {journal} {\bibinfo  {journal} {Nanotechnology}\ }\textbf {\bibinfo {volume} {32}},\ \bibinfo {pages} {065401} (\bibinfo {year} {2020}{\natexlab{a}})}\BibitemShut {NoStop}%
\bibitem [{\citenamefont {Felix}\ and\ \citenamefont {Pereira}(2020)}]{felix2020suppression}%
  \BibitemOpen
  \bibfield  {author} {\bibinfo {author} {\bibfnamefont {I.~M.}\ \bibnamefont {Felix}}\ and\ \bibinfo {author} {\bibfnamefont {L.~F.~C.}\ \bibnamefont {Pereira}},\ }\bibfield  {title} {\bibinfo {title} {Suppression of coherent thermal transport in quasiperiodic graphene-hbn superlattice ribbons},\ }\href@noop {} {\bibfield  {journal} {\bibinfo  {journal} {Carbon}\ }\textbf {\bibinfo {volume} {160}},\ \bibinfo {pages} {335} (\bibinfo {year} {2020})}\BibitemShut {NoStop}%
\bibitem [{\citenamefont {Yang}\ and\ \citenamefont {Chen}(2003)}]{yang2003partially}%
  \BibitemOpen
  \bibfield  {author} {\bibinfo {author} {\bibfnamefont {B.}~\bibnamefont {Yang}}\ and\ \bibinfo {author} {\bibfnamefont {G.}~\bibnamefont {Chen}},\ }\bibfield  {title} {\bibinfo {title} {Partially coherent phonon heat conduction in superlattices},\ }\href@noop {} {\bibfield  {journal} {\bibinfo  {journal} {Physical Review B}\ }\textbf {\bibinfo {volume} {67}},\ \bibinfo {pages} {195311} (\bibinfo {year} {2003})}\BibitemShut {NoStop}%
\bibitem [{\citenamefont {Ma}\ and\ \citenamefont {Wang}(2022)}]{ma2022ex}%
  \BibitemOpen
  \bibfield  {author} {\bibinfo {author} {\bibfnamefont {T.}~\bibnamefont {Ma}}\ and\ \bibinfo {author} {\bibfnamefont {Y.}~\bibnamefont {Wang}},\ }\bibfield  {title} {\bibinfo {title} {Ex-situ modification of lattice thermal transport through coherent and incoherent heat baths},\ }\href@noop {} {\bibfield  {journal} {\bibinfo  {journal} {Materials Today Physics}\ }\textbf {\bibinfo {volume} {29}},\ \bibinfo {pages} {100884} (\bibinfo {year} {2022})}\BibitemShut {NoStop}%
\bibitem [{\citenamefont {Qiu}\ \emph {et~al.}(2015)\citenamefont {Qiu}, \citenamefont {Chen},\ and\ \citenamefont {Tian}}]{qiu2015effects}%
  \BibitemOpen
  \bibfield  {author} {\bibinfo {author} {\bibfnamefont {B.}~\bibnamefont {Qiu}}, \bibinfo {author} {\bibfnamefont {G.}~\bibnamefont {Chen}},\ and\ \bibinfo {author} {\bibfnamefont {Z.}~\bibnamefont {Tian}},\ }\bibfield  {title} {\bibinfo {title} {Effects of aperiodicity and roughness on coherent heat conduction in superlattices},\ }\href@noop {} {\bibfield  {journal} {\bibinfo  {journal} {Nanoscale and Microscale Thermophysical Engineering}\ }\textbf {\bibinfo {volume} {19}},\ \bibinfo {pages} {272} (\bibinfo {year} {2015})}\BibitemShut {NoStop}%
\bibitem [{\citenamefont {Chakraborty}\ \emph {et~al.}(2017)\citenamefont {Chakraborty}, \citenamefont {Cao},\ and\ \citenamefont {Wang}}]{chakraborty2017ultralow}%
  \BibitemOpen
  \bibfield  {author} {\bibinfo {author} {\bibfnamefont {P.}~\bibnamefont {Chakraborty}}, \bibinfo {author} {\bibfnamefont {L.}~\bibnamefont {Cao}},\ and\ \bibinfo {author} {\bibfnamefont {Y.}~\bibnamefont {Wang}},\ }\bibfield  {title} {\bibinfo {title} {Ultralow lattice thermal conductivity of the random multilayer structure with lattice imperfections},\ }\href@noop {} {\bibfield  {journal} {\bibinfo  {journal} {Scientific reports}\ }\textbf {\bibinfo {volume} {7}},\ \bibinfo {pages} {8134} (\bibinfo {year} {2017})}\BibitemShut {NoStop}%
\bibitem [{\citenamefont {Chakraborty}\ \emph {et~al.}(2020{\natexlab{b}})\citenamefont {Chakraborty}, \citenamefont {Liu}, \citenamefont {Ma}, \citenamefont {Guo}, \citenamefont {Cao}, \citenamefont {Hu},\ and\ \citenamefont {Wang}}]{chakraborty2020quenching}%
  \BibitemOpen
  \bibfield  {author} {\bibinfo {author} {\bibfnamefont {P.}~\bibnamefont {Chakraborty}}, \bibinfo {author} {\bibfnamefont {Y.}~\bibnamefont {Liu}}, \bibinfo {author} {\bibfnamefont {T.}~\bibnamefont {Ma}}, \bibinfo {author} {\bibfnamefont {X.}~\bibnamefont {Guo}}, \bibinfo {author} {\bibfnamefont {L.}~\bibnamefont {Cao}}, \bibinfo {author} {\bibfnamefont {R.}~\bibnamefont {Hu}},\ and\ \bibinfo {author} {\bibfnamefont {Y.}~\bibnamefont {Wang}},\ }\bibfield  {title} {\bibinfo {title} {Quenching thermal transport in aperiodic superlattices: A molecular dynamics and machine learning study},\ }\href@noop {} {\bibfield  {journal} {\bibinfo  {journal} {ACS applied materials \& interfaces}\ }\textbf {\bibinfo {volume} {12}},\ \bibinfo {pages} {8795} (\bibinfo {year} {2020}{\natexlab{b}})}\BibitemShut {NoStop}%
\bibitem [{\citenamefont {Luo}\ and\ \citenamefont {Chen}(2013)}]{luo2013nanoscale}%
  \BibitemOpen
  \bibfield  {author} {\bibinfo {author} {\bibfnamefont {T.}~\bibnamefont {Luo}}\ and\ \bibinfo {author} {\bibfnamefont {G.}~\bibnamefont {Chen}},\ }\bibfield  {title} {\bibinfo {title} {Nanoscale heat transfer--from computation to experiment},\ }\href@noop {} {\bibfield  {journal} {\bibinfo  {journal} {Physical Chemistry Chemical Physics}\ }\textbf {\bibinfo {volume} {15}},\ \bibinfo {pages} {3389} (\bibinfo {year} {2013})}\BibitemShut {NoStop}%
\bibitem [{\citenamefont {Minnich}(2015)}]{minnich2015advances}%
  \BibitemOpen
  \bibfield  {author} {\bibinfo {author} {\bibfnamefont {A.}~\bibnamefont {Minnich}},\ }\bibfield  {title} {\bibinfo {title} {Advances in the measurement and computation of thermal phonon transport properties},\ }\href@noop {} {\bibfield  {journal} {\bibinfo  {journal} {Journal of Physics: Condensed Matter}\ }\textbf {\bibinfo {volume} {27}},\ \bibinfo {pages} {053202} (\bibinfo {year} {2015})}\BibitemShut {NoStop}%
\bibitem [{\citenamefont {Lindsay}\ \emph {et~al.}(2019)\citenamefont {Lindsay}, \citenamefont {Katre}, \citenamefont {Cepellotti},\ and\ \citenamefont {Mingo}}]{lindsay2019perspective}%
  \BibitemOpen
  \bibfield  {author} {\bibinfo {author} {\bibfnamefont {L.}~\bibnamefont {Lindsay}}, \bibinfo {author} {\bibfnamefont {A.}~\bibnamefont {Katre}}, \bibinfo {author} {\bibfnamefont {A.}~\bibnamefont {Cepellotti}},\ and\ \bibinfo {author} {\bibfnamefont {N.}~\bibnamefont {Mingo}},\ }\bibfield  {title} {\bibinfo {title} {Perspective on ab initio phonon thermal transport},\ }\href@noop {} {\bibfield  {journal} {\bibinfo  {journal} {Journal of Applied Physics}\ }\textbf {\bibinfo {volume} {126}} (\bibinfo {year} {2019})}\BibitemShut {NoStop}%
\bibitem [{\citenamefont {Schelling}\ \emph {et~al.}(2002)\citenamefont {Schelling}, \citenamefont {Phillpot},\ and\ \citenamefont {Keblinski}}]{schelling2002phonon}%
  \BibitemOpen
  \bibfield  {author} {\bibinfo {author} {\bibfnamefont {P.~K.}\ \bibnamefont {Schelling}}, \bibinfo {author} {\bibfnamefont {S.~R.}\ \bibnamefont {Phillpot}},\ and\ \bibinfo {author} {\bibfnamefont {P.}~\bibnamefont {Keblinski}},\ }\bibfield  {title} {\bibinfo {title} {Phonon wave-packet dynamics at semiconductor interfaces by molecular-dynamics simulation},\ }\href@noop {} {\bibfield  {journal} {\bibinfo  {journal} {Applied Physics Letters}\ }\textbf {\bibinfo {volume} {80}},\ \bibinfo {pages} {2484} (\bibinfo {year} {2002})}\BibitemShut {NoStop}%
\bibitem [{\citenamefont {Schelling}\ and\ \citenamefont {Phillpot}(2003)}]{schelling2003multiscale}%
  \BibitemOpen
  \bibfield  {author} {\bibinfo {author} {\bibfnamefont {P.}~\bibnamefont {Schelling}}\ and\ \bibinfo {author} {\bibfnamefont {S.}~\bibnamefont {Phillpot}},\ }\bibfield  {title} {\bibinfo {title} {Multiscale simulation of phonon transport in superlattices},\ }\href@noop {} {\bibfield  {journal} {\bibinfo  {journal} {Journal of Applied Physics}\ }\textbf {\bibinfo {volume} {93}},\ \bibinfo {pages} {5377} (\bibinfo {year} {2003})}\BibitemShut {NoStop}%
\bibitem [{\citenamefont {Tian}\ \emph {et~al.}(2010)\citenamefont {Tian}, \citenamefont {White},\ and\ \citenamefont {Sun}}]{tian2010phonon}%
  \BibitemOpen
  \bibfield  {author} {\bibinfo {author} {\bibfnamefont {Z.}~\bibnamefont {Tian}}, \bibinfo {author} {\bibfnamefont {B.}~\bibnamefont {White}},\ and\ \bibinfo {author} {\bibfnamefont {Y.}~\bibnamefont {Sun}},\ }\bibfield  {title} {\bibinfo {title} {Phonon wave-packet interference and phonon tunneling based energy transport across nanostructured thin films},\ }\href@noop {} {\bibfield  {journal} {\bibinfo  {journal} {Applied Physics Letters}\ }\textbf {\bibinfo {volume} {96}} (\bibinfo {year} {2010})}\BibitemShut {NoStop}%
\bibitem [{\citenamefont {Liang}\ \emph {et~al.}(2017)\citenamefont {Liang}, \citenamefont {Wilson},\ and\ \citenamefont {Keblinski}}]{liang2017phonon}%
  \BibitemOpen
  \bibfield  {author} {\bibinfo {author} {\bibfnamefont {Z.}~\bibnamefont {Liang}}, \bibinfo {author} {\bibfnamefont {T.~E.}\ \bibnamefont {Wilson}},\ and\ \bibinfo {author} {\bibfnamefont {P.}~\bibnamefont {Keblinski}},\ }\bibfield  {title} {\bibinfo {title} {Phonon interference in crystalline and amorphous confined nanoscopic films},\ }\href@noop {} {\bibfield  {journal} {\bibinfo  {journal} {Journal of Applied Physics}\ }\textbf {\bibinfo {volume} {121}} (\bibinfo {year} {2017})}\BibitemShut {NoStop}%
\bibitem [{\citenamefont {Shao}\ \emph {et~al.}(2018)\citenamefont {Shao}, \citenamefont {Rong}, \citenamefont {Li},\ and\ \citenamefont {Bao}}]{shao2018understanding}%
  \BibitemOpen
  \bibfield  {author} {\bibinfo {author} {\bibfnamefont {C.}~\bibnamefont {Shao}}, \bibinfo {author} {\bibfnamefont {Q.}~\bibnamefont {Rong}}, \bibinfo {author} {\bibfnamefont {N.}~\bibnamefont {Li}},\ and\ \bibinfo {author} {\bibfnamefont {H.}~\bibnamefont {Bao}},\ }\bibfield  {title} {\bibinfo {title} {Understanding the mechanism of diffuse phonon scattering at disordered surfaces by atomistic wave-packet investigation},\ }\href@noop {} {\bibfield  {journal} {\bibinfo  {journal} {Physical Review B}\ }\textbf {\bibinfo {volume} {98}},\ \bibinfo {pages} {155418} (\bibinfo {year} {2018})}\BibitemShut {NoStop}%
\bibitem [{\citenamefont {Shi}\ \emph {et~al.}(2018)\citenamefont {Shi}, \citenamefont {Lee}, \citenamefont {Dong}, \citenamefont {Roy}, \citenamefont {Fisher},\ and\ \citenamefont {Ruan}}]{shi2018dominant}%
  \BibitemOpen
  \bibfield  {author} {\bibinfo {author} {\bibfnamefont {J.}~\bibnamefont {Shi}}, \bibinfo {author} {\bibfnamefont {J.}~\bibnamefont {Lee}}, \bibinfo {author} {\bibfnamefont {Y.}~\bibnamefont {Dong}}, \bibinfo {author} {\bibfnamefont {A.}~\bibnamefont {Roy}}, \bibinfo {author} {\bibfnamefont {T.~S.}\ \bibnamefont {Fisher}},\ and\ \bibinfo {author} {\bibfnamefont {X.}~\bibnamefont {Ruan}},\ }\bibfield  {title} {\bibinfo {title} {Dominant phonon polarization conversion across dimensionally mismatched interfaces: Carbon-nanotube--graphene junction},\ }\href@noop {} {\bibfield  {journal} {\bibinfo  {journal} {Physical Review B}\ }\textbf {\bibinfo {volume} {97}},\ \bibinfo {pages} {134309} (\bibinfo {year} {2018})}\BibitemShut {NoStop}%
\bibitem [{\citenamefont {Desmarchelier}\ \emph {et~al.}(2021)\citenamefont {Desmarchelier}, \citenamefont {Carr{\'e}}, \citenamefont {Termentzidis},\ and\ \citenamefont {Tanguy}}]{desmarchelier2021ballistic}%
  \BibitemOpen
  \bibfield  {author} {\bibinfo {author} {\bibfnamefont {P.}~\bibnamefont {Desmarchelier}}, \bibinfo {author} {\bibfnamefont {A.}~\bibnamefont {Carr{\'e}}}, \bibinfo {author} {\bibfnamefont {K.}~\bibnamefont {Termentzidis}},\ and\ \bibinfo {author} {\bibfnamefont {A.}~\bibnamefont {Tanguy}},\ }\bibfield  {title} {\bibinfo {title} {Ballistic heat transport in nanocomposite: The role of the shape and interconnection of nanoinclusions},\ }\href@noop {} {\bibfield  {journal} {\bibinfo  {journal} {Nanomaterials}\ }\textbf {\bibinfo {volume} {11}},\ \bibinfo {pages} {1982} (\bibinfo {year} {2021})}\BibitemShut {NoStop}%
\bibitem [{\citenamefont {Jiang}\ \emph {et~al.}(2021)\citenamefont {Jiang}, \citenamefont {Ouyang}, \citenamefont {Ren}, \citenamefont {Yu}, \citenamefont {He},\ and\ \citenamefont {Chen}}]{jiang2021total}%
  \BibitemOpen
  \bibfield  {author} {\bibinfo {author} {\bibfnamefont {P.}~\bibnamefont {Jiang}}, \bibinfo {author} {\bibfnamefont {Y.}~\bibnamefont {Ouyang}}, \bibinfo {author} {\bibfnamefont {W.}~\bibnamefont {Ren}}, \bibinfo {author} {\bibfnamefont {C.}~\bibnamefont {Yu}}, \bibinfo {author} {\bibfnamefont {J.}~\bibnamefont {He}},\ and\ \bibinfo {author} {\bibfnamefont {J.}~\bibnamefont {Chen}},\ }\bibfield  {title} {\bibinfo {title} {Total-transmission and total-reflection of individual phonons in phononic crystal nanostructures},\ }\href@noop {} {\bibfield  {journal} {\bibinfo  {journal} {APL Materials}\ }\textbf {\bibinfo {volume} {9}} (\bibinfo {year} {2021})}\BibitemShut {NoStop}%
\bibitem [{\citenamefont {Maranets}\ and\ \citenamefont {Wang}(2023)}]{maranets2023ballistic}%
  \BibitemOpen
  \bibfield  {author} {\bibinfo {author} {\bibfnamefont {T.}~\bibnamefont {Maranets}}\ and\ \bibinfo {author} {\bibfnamefont {Y.}~\bibnamefont {Wang}},\ }\bibfield  {title} {\bibinfo {title} {Ballistic phonon lensing by the non-planar interfaces of embedded nanoparticles},\ }\href@noop {} {\bibfield  {journal} {\bibinfo  {journal} {New Journal of Physics}\ }\textbf {\bibinfo {volume} {25}},\ \bibinfo {pages} {103038} (\bibinfo {year} {2023})}\BibitemShut {NoStop}%
\bibitem [{\citenamefont {Beardo}\ \emph {et~al.}(2024)\citenamefont {Beardo}, \citenamefont {Desmarchelier}, \citenamefont {Tsai}, \citenamefont {Rawte}, \citenamefont {Termentzidis},\ and\ \citenamefont {Hussein}}]{beardo2024resonant}%
  \BibitemOpen
  \bibfield  {author} {\bibinfo {author} {\bibfnamefont {A.}~\bibnamefont {Beardo}}, \bibinfo {author} {\bibfnamefont {P.}~\bibnamefont {Desmarchelier}}, \bibinfo {author} {\bibfnamefont {C.-N.}\ \bibnamefont {Tsai}}, \bibinfo {author} {\bibfnamefont {P.}~\bibnamefont {Rawte}}, \bibinfo {author} {\bibfnamefont {K.}~\bibnamefont {Termentzidis}},\ and\ \bibinfo {author} {\bibfnamefont {M.~I.}\ \bibnamefont {Hussein}},\ }\bibfield  {title} {\bibinfo {title} {Resonant phonons: Localization in a structurally ordered crystal},\ }\href@noop {} {\bibfield  {journal} {\bibinfo  {journal} {Physical Review B}\ }\textbf {\bibinfo {volume} {110}},\ \bibinfo {pages} {195438} (\bibinfo {year} {2024})}\BibitemShut {NoStop}%
\bibitem [{\citenamefont {Thompson}\ \emph {et~al.}(2022)\citenamefont {Thompson}, \citenamefont {Aktulga}, \citenamefont {Berger}, \citenamefont {Bolintineanu}, \citenamefont {Brown}, \citenamefont {Crozier}, \citenamefont {In't~Veld}, \citenamefont {Kohlmeyer}, \citenamefont {Moore}, \citenamefont {Nguyen} \emph {et~al.}}]{thompson2022lammps}%
  \BibitemOpen
  \bibfield  {author} {\bibinfo {author} {\bibfnamefont {A.~P.}\ \bibnamefont {Thompson}}, \bibinfo {author} {\bibfnamefont {H.~M.}\ \bibnamefont {Aktulga}}, \bibinfo {author} {\bibfnamefont {R.}~\bibnamefont {Berger}}, \bibinfo {author} {\bibfnamefont {D.~S.}\ \bibnamefont {Bolintineanu}}, \bibinfo {author} {\bibfnamefont {W.~M.}\ \bibnamefont {Brown}}, \bibinfo {author} {\bibfnamefont {P.~S.}\ \bibnamefont {Crozier}}, \bibinfo {author} {\bibfnamefont {P.~J.}\ \bibnamefont {In't~Veld}}, \bibinfo {author} {\bibfnamefont {A.}~\bibnamefont {Kohlmeyer}}, \bibinfo {author} {\bibfnamefont {S.~G.}\ \bibnamefont {Moore}}, \bibinfo {author} {\bibfnamefont {T.~D.}\ \bibnamefont {Nguyen}}, \emph {et~al.},\ }\bibfield  {title} {\bibinfo {title} {Lammps-a flexible simulation tool for particle-based materials modeling at the atomic, meso, and continuum scales},\ }\href@noop {} {\bibfield  {journal} {\bibinfo  {journal} {Computer Physics Communications}\ }\textbf {\bibinfo {volume} {271}},\ \bibinfo {pages} {108171}
  (\bibinfo {year} {2022})}\BibitemShut {NoStop}%
\bibitem [{\citenamefont {Tamura}\ \emph {et~al.}(1988)\citenamefont {Tamura}, \citenamefont {Hurley},\ and\ \citenamefont {Wolfe}}]{tamura1988acoustic}%
  \BibitemOpen
  \bibfield  {author} {\bibinfo {author} {\bibfnamefont {S.}~\bibnamefont {Tamura}}, \bibinfo {author} {\bibfnamefont {D.}~\bibnamefont {Hurley}},\ and\ \bibinfo {author} {\bibfnamefont {J.}~\bibnamefont {Wolfe}},\ }\bibfield  {title} {\bibinfo {title} {Acoustic-phonon propagation in superlattices},\ }\href@noop {} {\bibfield  {journal} {\bibinfo  {journal} {Physical Review B}\ }\textbf {\bibinfo {volume} {38}},\ \bibinfo {pages} {1427} (\bibinfo {year} {1988})}\BibitemShut {NoStop}%
\bibitem [{\citenamefont {Tamura}\ and\ \citenamefont {Wolfe}(1988)}]{tamura1988acousticmulti}%
  \BibitemOpen
  \bibfield  {author} {\bibinfo {author} {\bibfnamefont {S.}~\bibnamefont {Tamura}}\ and\ \bibinfo {author} {\bibfnamefont {J.}~\bibnamefont {Wolfe}},\ }\bibfield  {title} {\bibinfo {title} {Acoustic phonons in multiconstituent superlattices},\ }\href@noop {} {\bibfield  {journal} {\bibinfo  {journal} {Physical Review B}\ }\textbf {\bibinfo {volume} {38}},\ \bibinfo {pages} {5610} (\bibinfo {year} {1988})}\BibitemShut {NoStop}%
\bibitem [{\citenamefont {Tamura}(1989)}]{tamura1989localized}%
  \BibitemOpen
  \bibfield  {author} {\bibinfo {author} {\bibfnamefont {S.-i.}\ \bibnamefont {Tamura}},\ }\bibfield  {title} {\bibinfo {title} {Localized vibrational modes in superlattices},\ }\href@noop {} {\bibfield  {journal} {\bibinfo  {journal} {Physical Review B}\ }\textbf {\bibinfo {volume} {39}},\ \bibinfo {pages} {1261} (\bibinfo {year} {1989})}\BibitemShut {NoStop}%
\bibitem [{\citenamefont {Hurley}\ \emph {et~al.}(1987)\citenamefont {Hurley}, \citenamefont {Tamura}, \citenamefont {Wolfe},\ and\ \citenamefont {Morkoc}}]{hurley1987imaging}%
  \BibitemOpen
  \bibfield  {author} {\bibinfo {author} {\bibfnamefont {D.}~\bibnamefont {Hurley}}, \bibinfo {author} {\bibfnamefont {S.}~\bibnamefont {Tamura}}, \bibinfo {author} {\bibfnamefont {J.}~\bibnamefont {Wolfe}},\ and\ \bibinfo {author} {\bibfnamefont {H.}~\bibnamefont {Morkoc}},\ }\bibfield  {title} {\bibinfo {title} {Imaging of acoustic phonon stop bands in superlattices},\ }\href@noop {} {\bibfield  {journal} {\bibinfo  {journal} {Physical review letters}\ }\textbf {\bibinfo {volume} {58}},\ \bibinfo {pages} {2446} (\bibinfo {year} {1987})}\BibitemShut {NoStop}%
\bibitem [{\citenamefont {Tanaka}\ \emph {et~al.}(1998)\citenamefont {Tanaka}, \citenamefont {Narita},\ and\ \citenamefont {Tamura}}]{tanaka1998phonon}%
  \BibitemOpen
  \bibfield  {author} {\bibinfo {author} {\bibfnamefont {Y.}~\bibnamefont {Tanaka}}, \bibinfo {author} {\bibfnamefont {M.}~\bibnamefont {Narita}},\ and\ \bibinfo {author} {\bibfnamefont {S.-i.}\ \bibnamefont {Tamura}},\ }\bibfield  {title} {\bibinfo {title} {Phonon focusing in superlattices},\ }\href@noop {} {\bibfield  {journal} {\bibinfo  {journal} {Journal of Physics: Condensed Matter}\ }\textbf {\bibinfo {volume} {10}},\ \bibinfo {pages} {8787} (\bibinfo {year} {1998})}\BibitemShut {NoStop}%
\bibitem [{\citenamefont {Baker}\ \emph {et~al.}(2012)\citenamefont {Baker}, \citenamefont {Jordan},\ and\ \citenamefont {Norris}}]{baker2012application}%
  \BibitemOpen
  \bibfield  {author} {\bibinfo {author} {\bibfnamefont {C.~H.}\ \bibnamefont {Baker}}, \bibinfo {author} {\bibfnamefont {D.~A.}\ \bibnamefont {Jordan}},\ and\ \bibinfo {author} {\bibfnamefont {P.~M.}\ \bibnamefont {Norris}},\ }\bibfield  {title} {\bibinfo {title} {Application of the wavelet transform to nanoscale thermal transport},\ }\href {https://doi.org/10.1103/PhysRevB.86.104306} {\bibfield  {journal} {\bibinfo  {journal} {Phys. Rev. B}\ }\textbf {\bibinfo {volume} {86}},\ \bibinfo {pages} {104306} (\bibinfo {year} {2012})}\BibitemShut {NoStop}%
\bibitem [{\citenamefont {Thomas}\ \emph {et~al.}(2010)\citenamefont {Thomas}, \citenamefont {Turney}, \citenamefont {Iutzi}, \citenamefont {Amon},\ and\ \citenamefont {McGaughey}}]{thomas2010predicting}%
  \BibitemOpen
  \bibfield  {author} {\bibinfo {author} {\bibfnamefont {J.~A.}\ \bibnamefont {Thomas}}, \bibinfo {author} {\bibfnamefont {J.~E.}\ \bibnamefont {Turney}}, \bibinfo {author} {\bibfnamefont {R.~M.}\ \bibnamefont {Iutzi}}, \bibinfo {author} {\bibfnamefont {C.~H.}\ \bibnamefont {Amon}},\ and\ \bibinfo {author} {\bibfnamefont {A.~J.}\ \bibnamefont {McGaughey}},\ }\bibfield  {title} {\bibinfo {title} {Predicting phonon dispersion relations and lifetimes from the spectral energy density},\ }\href@noop {} {\bibfield  {journal} {\bibinfo  {journal} {Physical Review B—Condensed Matter and Materials Physics}\ }\textbf {\bibinfo {volume} {81}},\ \bibinfo {pages} {081411} (\bibinfo {year} {2010})}\BibitemShut {NoStop}%
\bibitem [{\citenamefont {Tamura}\ and\ \citenamefont {Nori}(1990)}]{tamura1990acoustic}%
  \BibitemOpen
  \bibfield  {author} {\bibinfo {author} {\bibfnamefont {S.-i.}\ \bibnamefont {Tamura}}\ and\ \bibinfo {author} {\bibfnamefont {F.}~\bibnamefont {Nori}},\ }\bibfield  {title} {\bibinfo {title} {Acoustic interference in random superlattices},\ }\href@noop {} {\bibfield  {journal} {\bibinfo  {journal} {Physical Review B}\ }\textbf {\bibinfo {volume} {41}},\ \bibinfo {pages} {7941} (\bibinfo {year} {1990})}\BibitemShut {NoStop}%
\bibitem [{\citenamefont {Allen}\ and\ \citenamefont {Feldman}(1993)}]{allen1993thermal}%
  \BibitemOpen
  \bibfield  {author} {\bibinfo {author} {\bibfnamefont {P.~B.}\ \bibnamefont {Allen}}\ and\ \bibinfo {author} {\bibfnamefont {J.~L.}\ \bibnamefont {Feldman}},\ }\bibfield  {title} {\bibinfo {title} {Thermal conductivity of disordered harmonic solids},\ }\href@noop {} {\bibfield  {journal} {\bibinfo  {journal} {Physical Review B}\ }\textbf {\bibinfo {volume} {48}},\ \bibinfo {pages} {12581} (\bibinfo {year} {1993})}\BibitemShut {NoStop}%
\bibitem [{\citenamefont {Feldman}\ \emph {et~al.}(1993)\citenamefont {Feldman}, \citenamefont {Kluge}, \citenamefont {Allen},\ and\ \citenamefont {Wooten}}]{feldman1993thermal}%
  \BibitemOpen
  \bibfield  {author} {\bibinfo {author} {\bibfnamefont {J.~L.}\ \bibnamefont {Feldman}}, \bibinfo {author} {\bibfnamefont {M.~D.}\ \bibnamefont {Kluge}}, \bibinfo {author} {\bibfnamefont {P.~B.}\ \bibnamefont {Allen}},\ and\ \bibinfo {author} {\bibfnamefont {F.}~\bibnamefont {Wooten}},\ }\bibfield  {title} {\bibinfo {title} {Thermal conductivity and localization in glasses: Numerical study of a model of amorphous silicon},\ }\href@noop {} {\bibfield  {journal} {\bibinfo  {journal} {Physical Review B}\ }\textbf {\bibinfo {volume} {48}},\ \bibinfo {pages} {12589} (\bibinfo {year} {1993})}\BibitemShut {NoStop}%
\bibitem [{\citenamefont {Allen}\ \emph {et~al.}(1999)\citenamefont {Allen}, \citenamefont {Feldman}, \citenamefont {Fabian},\ and\ \citenamefont {Wooten}}]{allen1999diffusons}%
  \BibitemOpen
  \bibfield  {author} {\bibinfo {author} {\bibfnamefont {P.~B.}\ \bibnamefont {Allen}}, \bibinfo {author} {\bibfnamefont {J.~L.}\ \bibnamefont {Feldman}}, \bibinfo {author} {\bibfnamefont {J.}~\bibnamefont {Fabian}},\ and\ \bibinfo {author} {\bibfnamefont {F.}~\bibnamefont {Wooten}},\ }\bibfield  {title} {\bibinfo {title} {Diffusons, locons and propagons: Character of atomic vibrations in amorphous si},\ }\href@noop {} {\bibfield  {journal} {\bibinfo  {journal} {Philosophical Magazine B}\ }\textbf {\bibinfo {volume} {79}},\ \bibinfo {pages} {1715} (\bibinfo {year} {1999})}\BibitemShut {NoStop}%
\bibitem [{\citenamefont {Lv}\ and\ \citenamefont {Henry}(2016)}]{lv2016examining}%
  \BibitemOpen
  \bibfield  {author} {\bibinfo {author} {\bibfnamefont {W.}~\bibnamefont {Lv}}\ and\ \bibinfo {author} {\bibfnamefont {A.}~\bibnamefont {Henry}},\ }\bibfield  {title} {\bibinfo {title} {Examining the validity of the phonon gas model in amorphous materials},\ }\href@noop {} {\bibfield  {journal} {\bibinfo  {journal} {Scientific reports}\ }\textbf {\bibinfo {volume} {6}},\ \bibinfo {pages} {37675} (\bibinfo {year} {2016})}\BibitemShut {NoStop}%
\bibitem [{\citenamefont {Seyf}\ \emph {et~al.}(2017)\citenamefont {Seyf}, \citenamefont {Yates}, \citenamefont {Bougher}, \citenamefont {Graham}, \citenamefont {Cola}, \citenamefont {Detchprohm}, \citenamefont {Ji}, \citenamefont {Kim}, \citenamefont {Dupuis}, \citenamefont {Lv} \emph {et~al.}}]{seyf2017rethinking}%
  \BibitemOpen
  \bibfield  {author} {\bibinfo {author} {\bibfnamefont {H.~R.}\ \bibnamefont {Seyf}}, \bibinfo {author} {\bibfnamefont {L.}~\bibnamefont {Yates}}, \bibinfo {author} {\bibfnamefont {T.~L.}\ \bibnamefont {Bougher}}, \bibinfo {author} {\bibfnamefont {S.}~\bibnamefont {Graham}}, \bibinfo {author} {\bibfnamefont {B.~A.}\ \bibnamefont {Cola}}, \bibinfo {author} {\bibfnamefont {T.}~\bibnamefont {Detchprohm}}, \bibinfo {author} {\bibfnamefont {M.-H.}\ \bibnamefont {Ji}}, \bibinfo {author} {\bibfnamefont {J.}~\bibnamefont {Kim}}, \bibinfo {author} {\bibfnamefont {R.}~\bibnamefont {Dupuis}}, \bibinfo {author} {\bibfnamefont {W.}~\bibnamefont {Lv}}, \emph {et~al.},\ }\bibfield  {title} {\bibinfo {title} {Rethinking phonons: The issue of disorder},\ }\href@noop {} {\bibfield  {journal} {\bibinfo  {journal} {npj Computational Materials}\ }\textbf {\bibinfo {volume} {3}},\ \bibinfo {pages} {49} (\bibinfo {year} {2017})}\BibitemShut {NoStop}%
\bibitem [{\citenamefont {Simoncelli}\ \emph {et~al.}(2019)\citenamefont {Simoncelli}, \citenamefont {Marzari},\ and\ \citenamefont {Mauri}}]{simoncelli2019unified}%
  \BibitemOpen
  \bibfield  {author} {\bibinfo {author} {\bibfnamefont {M.}~\bibnamefont {Simoncelli}}, \bibinfo {author} {\bibfnamefont {N.}~\bibnamefont {Marzari}},\ and\ \bibinfo {author} {\bibfnamefont {F.}~\bibnamefont {Mauri}},\ }\bibfield  {title} {\bibinfo {title} {Unified theory of thermal transport in crystals and glasses},\ }\href@noop {} {\bibfield  {journal} {\bibinfo  {journal} {Nature Physics}\ }\textbf {\bibinfo {volume} {15}},\ \bibinfo {pages} {809} (\bibinfo {year} {2019})}\BibitemShut {NoStop}%
\bibitem [{\citenamefont {Isaeva}\ \emph {et~al.}(2019)\citenamefont {Isaeva}, \citenamefont {Barbalinardo}, \citenamefont {Donadio},\ and\ \citenamefont {Baroni}}]{isaeva2019modeling}%
  \BibitemOpen
  \bibfield  {author} {\bibinfo {author} {\bibfnamefont {L.}~\bibnamefont {Isaeva}}, \bibinfo {author} {\bibfnamefont {G.}~\bibnamefont {Barbalinardo}}, \bibinfo {author} {\bibfnamefont {D.}~\bibnamefont {Donadio}},\ and\ \bibinfo {author} {\bibfnamefont {S.}~\bibnamefont {Baroni}},\ }\bibfield  {title} {\bibinfo {title} {Modeling heat transport in crystals and glasses from a unified lattice-dynamical approach},\ }\href@noop {} {\bibfield  {journal} {\bibinfo  {journal} {Nature communications}\ }\textbf {\bibinfo {volume} {10}},\ \bibinfo {pages} {3853} (\bibinfo {year} {2019})}\BibitemShut {NoStop}%
\bibitem [{\citenamefont {Bodapati}\ \emph {et~al.}(2006)\citenamefont {Bodapati}, \citenamefont {Schelling}, \citenamefont {Phillpot},\ and\ \citenamefont {Keblinski}}]{bodapati2006vibrations}%
  \BibitemOpen
  \bibfield  {author} {\bibinfo {author} {\bibfnamefont {A.}~\bibnamefont {Bodapati}}, \bibinfo {author} {\bibfnamefont {P.~K.}\ \bibnamefont {Schelling}}, \bibinfo {author} {\bibfnamefont {S.~R.}\ \bibnamefont {Phillpot}},\ and\ \bibinfo {author} {\bibfnamefont {P.}~\bibnamefont {Keblinski}},\ }\bibfield  {title} {\bibinfo {title} {Vibrations and thermal transport in nanocrystalline silicon},\ }\href@noop {} {\bibfield  {journal} {\bibinfo  {journal} {Physical Review B—Condensed Matter and Materials Physics}\ }\textbf {\bibinfo {volume} {74}},\ \bibinfo {pages} {245207} (\bibinfo {year} {2006})}\BibitemShut {NoStop}%
\bibitem [{\citenamefont {Zhang}\ \emph {et~al.}(2022)\citenamefont {Zhang}, \citenamefont {Guo}, \citenamefont {Bescond}, \citenamefont {Chen}, \citenamefont {Nomura},\ and\ \citenamefont {Volz}}]{zhang2022coherence}%
  \BibitemOpen
  \bibfield  {author} {\bibinfo {author} {\bibfnamefont {Z.}~\bibnamefont {Zhang}}, \bibinfo {author} {\bibfnamefont {Y.}~\bibnamefont {Guo}}, \bibinfo {author} {\bibfnamefont {M.}~\bibnamefont {Bescond}}, \bibinfo {author} {\bibfnamefont {J.}~\bibnamefont {Chen}}, \bibinfo {author} {\bibfnamefont {M.}~\bibnamefont {Nomura}},\ and\ \bibinfo {author} {\bibfnamefont {S.}~\bibnamefont {Volz}},\ }\bibfield  {title} {\bibinfo {title} {How coherence is governing diffuson heat transfer in amorphous solids},\ }\href@noop {} {\bibfield  {journal} {\bibinfo  {journal} {npj Computational Materials}\ }\textbf {\bibinfo {volume} {8}},\ \bibinfo {pages} {96} (\bibinfo {year} {2022})}\BibitemShut {NoStop}%
\end{thebibliography}%

\end{document}